\documentclass[letterpaper,11pt]{article}
\pdfoutput=1 

\usepackage{jheppub} 
            
\usepackage{bm,amssymb,slashed,graphicx,multirow,soul,mathtools,xspace,array}  
\usepackage{float}                   
\allowdisplaybreaks
\usepackage{ bbold }
\usepackage{subfigure}

\newcolumntype{L}[1]{>{\arraybackslash}p{#1}}

\newcommand{\todo}[1]{{\color{red} \ifmmode\else[todo]\fi #1}}
\usepackage[usenames,dvipsnames]{xcolor}
     \definecolor{hgreen}{rgb}{0,.3,0}
      \definecolor{darkgreen}{rgb}{0.3,.8,0.2}
     \definecolor{hred}{rgb}{.3,0,0}
     \definecolor{hblue}{rgb}{0,0,.3}
     \definecolor{LightGray}{gray}{0.95}
     
\usepackage{hyperref}

\usepackage{mathrsfs}

 \usepackage{etoolbox}
    \makeatletter
    \patchcmd{\maketitle}{\@fpheader}{}{}{}
    \makeatother

\newcommand\snowmass{
\begin{center}
  \rule[-0.2in]{\hsize}{0.01in}\\
  \rule{\hsize}{0.01in}\\
  \vskip 0.1in
  Submitted to the Proceedings of the US Community Study\\ 
  on the Future of Particle Physics (Snowmass 2021)\\
  \rule{\hsize}{0.01in}\\
  \rule[+0.2in]{\hsize}{0.01in}\\[-2em]
\end{center}
}

\def\beq{\begin{equation}}
\def\eeq{\end{equation}}

\title{Snowmass White Paper: New flavors and rich structures in dark sectors}

\author[a]{Philip Harris,}
\author[b]{Philip Schuster,}
\author[c]{Jure Zupan}

\affiliation[a]{Massachusetts Institute of Technology, Cambridge, MA 02139, USA}
\affiliation[b]{SLAC National Accelerator Laboratory, Menlo Park, CA 94025, USA}
\affiliation[c]{Department of Physics, University of Cincinnati, Cincinnati, Ohio 45221, USA}

\date{\today}

\abstract{Dark matter can be part of a dark sector with non-minimal couplings to the Standard Model. Compared to many (minimal) benchmark models, such scenarios can result in significant modifications in experimental signatures and strongly impact experimental search sensitivity. In this white paper, we review several non-minimal dark sector models, including phenomenological consequences: models explaining $(g-2)_\mu$, inelastic dark matter, strongly interacting massive particles as dark matter candidates, and axions with flavorful couplings. The present exclusions and projected experimental sensitivities on these example dark sector models illustrate the robustness of the growing dark sector experimental effort -- both the broadness and the precision of existing searches -- probing theoretically interesting parameter space. They also illustrate some of the unique complementarity of different experimental approaches.}

\begin{document}

\snowmass

 \maketitle
 
 \vfill

\pagebreak
%


\section{Executive Summary}
Dark matter (DM) directly  motivates the existence of a dark sector of matter and interactions beyond the Standard Model. To-date, much of the emphasis for experimental work on dark sectors has been anchored to minimal models, often with only a single mediator particle, single dark matter candidate, and the assumption of flavor universality in the interactions. However, like the Standard Model, dark sectors may have non-minimal structures, either in couplings to the Standard Model, or in the spectra of dark sector states. Often this leads to a far richer phenomenology and may require rethinking of experimental strategies for  achieving optimized sensitivities. Additionally, extensions to the dark sector can help resolve important theoretical puzzles and data driven anomalies. The purpose of this white paper is to showcase examples where interesting non-minimal dark sector phenomena can be efficiently searched for and uncovered in high intensity experiments. The examples are organized into three themes -- examples motivated by data driven anomalies, by theoretical puzzles, and by more complete coverage of the  commonly used benchmark dark sector models, going beyond the assumption of minimality. 

From a theory point of view, dark sectors coupled to the Standard Model by means of the so-called ``portal'' operators \cite{Patt:2006fw,Beacham:2019nyx,Batell:2009di} represent a framework that encapsulates a broad range of new physics ideas while respecting the well-measured symmetries of the Standard Model. As such, it is not surprising that the framework of dark sector portals is widely used to study new physics explanations for a range of data-driven anomalies, such as those in the measurements of $(g-2)_\mu$ \cite{Muong-2:2021ojo,Aoyama:2020ynm,Muong-2:2006rrc}, rare $b\to c\tau \nu$ \cite{HFLAV:2019otj,Bernlochner:2021vlv,BaBar:2012obs,BaBar:2013mob,Belle:2015qfa,LHCb:2015gmp,Belle:2016dyj,LHCb:2017smo,LHCb:2017rln,Belle:2019rba} and $b\to s \mu^+\mu^-$ decays \cite{LHCb:2020lmf,LHCb:2020gog,LHCb:2021awg,LHCb:2021vsc,LHCb:2014cxe,LHCb:2015wdu,LHCb:2016ykl,LHCb:2021zwz,LHCb:2017avl,LHCb:2021trn,Altmannshofer:2021qrr,Geng:2021nhg,Alguero:2021anc,Hurth:2021nsi}, the Xenon 1T excess \cite{XENON:2020rca}, the MiniBoone excess \cite{MiniBooNE:2010idf,MiniBooNE:2013uba,MiniBooNE:2018esg,MiniBooNE:2020pnu}, the Beryllium excited state decays \cite{Krasznahorkay:2015iga,Feng:2016jff,Krasznahorkay:2019lyl,Krasznahorkay:2021joi}, the neutron lifetime puzzle \cite{Fornal:2020gto}, and KOTO anomalous events \cite{KOTO:2020prk,Goudzovski:2022vbt}, among others. Among these, $(g-2)_\mu$ is the  longest standing data driven anomaly. We feature below several new physics models that can explain the discrepancy between the measurements and the consensus Standard Model prediction. In the case of $(g-2)_\mu$ anomaly the program of dark sector experiments has already had considerable impact, ruling out the simplest new physics ideas that were put forward to address the anomaly. The parameter space that remains involves non-minimal, flavor dependent, interactions. High intensity dark sector experiments will be able to probe most of the remaining explanations in the coming years, either ruling out the new physics explanations or making discoveries. 

The dark sector framework includes generalizations of the long sought after QCD axion, $a$, parametrizing it as part of a more general class of pseudo-scalar dark sector particles, referred to as axion-like particles (ALPs). The high intensity dark sector experimental program is able to probe all ALP couplings to the SM particles, including flavor violating couplings to quarks and leptons, for a broad range of masses. In this way, one probes also the non-minimal QCD axion models, in many cases probing complementary parameter space to  the searches based on axion couplings to photons, with the reach above astrophysics constraints. We highlight this below for the case of flavor violating QCD axion model, where searches for $s\to d a$ and $\mu \to e a$ transitions can probe the parameter range in which QCD axion is a viable cold dark matter candidate. 

The range of more complete models that include the most common minimal dark sector benchmarks is rather large, see, e.g., Refs. \cite{Lin:2019uvt,Petraki:2013wwa,Battaglieri:2017aum,Feng:2010gw,Bertone:2004pz,Boehm:2003hm,Jungman:1995df}, which is part of the reason the minimal models receive more attention. A common element of the more complete models concerns the origin of the dark sector mass scale, with the models split roughly into two categories -- the weak and strong coupling regime. We give a representative example from each category, namely inelastic dark matter (iDM), and the strongly interacting massive particles (SIMPs). Many minimal dark sector dark matter benchmarks include a massive vector mediator and a scalar or fermion dark matter candidate. In UV completions of these scenarios, the ``dark higgs'' that gives mass to the vector can also  split the dark matter multiplet into a pair of nearly degenerate states, the lightest of which is stable. Similar phenomenology can be obtained in the second category of models, in which the vector mediator mass instead arises from strong dynamics. Strong dynamics naturally results in dark sector mesons, and thus also predicts a wide range of new phenomena with displaced decays of the mesons to final states with Standard Model leptons.  For both iDM and SIMPs, viable predictions of the relic density exist. However, the dynamics changes with respect to the portal benchmark models. SIMPs models utilize 3$\rightarrow$2 processes to produce dark matter in the early universe, whereas iDM and the portal models rely on dark matter pair annihilation (2$\rightarrow$2 processes). Despite these modifications, it is possible to obtain dark matter densities consistent with the observed cosmological bounds. Furthermore, experiments in the next 10 years will be able to probe a significant portion of the allowed parameter space that would lead to viable dark matter relic density.  

The above examples  illustrate the robustness of the growing dark sector experimental effort, both the broadness of the different searches as well as the precision, probing theoretically well motivated parameter space.



\section{Framing of the white paper within the RF6 topical group}
\label{sec:Framing}

The challenge of searching for dark matter is dependent on the possible final states that can describe dark matter. Dark matter is often characterized through a benchmark model consisting of two particles, a portal and a dark matter candidate. Three main features describe this model. These features consist of the mass of the portal particle, the mass of the dark matter, and the relative interaction strengths of the portal particle with the dark matter. By varying the model features, it is possible to construct a variety of scenarios that can both explain the formation of dark matter in the early universe and a means to observe dark matter at a collider or through indirect or direct dark matter detection.  

Dark sector signatures at colliders fall into two sets of final states, visible and invisible. These final states are often the driving motivators for new dark sector experiments for the benchmark models. The invisible final state covers the scenario when dark matter is lighter than half the portal mediator mass and, as a result, the portal mediator predominately decays invisibly. The visible final state covers the scenario when dark matter is heavy and, as a result, the mediator decays into standard model particles. The decays to visible final states are also present, if the mediator does not couple to dark matter, and thus only a mediator is directly accessible. Several variations of these final states are also possible. When the coupling for dark matter to matter is sufficiently large and a detector is sufficiently sensitive, it is possible to observe dark matter scattering. Furthermore, when the coupling of the portal is sufficiently small, and the dark matter is heavy, the portal can be long-lived, leading to standard model final states that are displaced. The variety of final states obtained motivates a range of possible detector designs aimed at probing the benchmark models. However, it may be the case that the dark sector models considered are more complicated than the minimal example of a single dark matter particle and a mediator, and, as a result, a number of different non-minimal experimental signatures can also be observed. 

\begin{figure}
    \centering
    \includegraphics[width=0.555\textwidth]{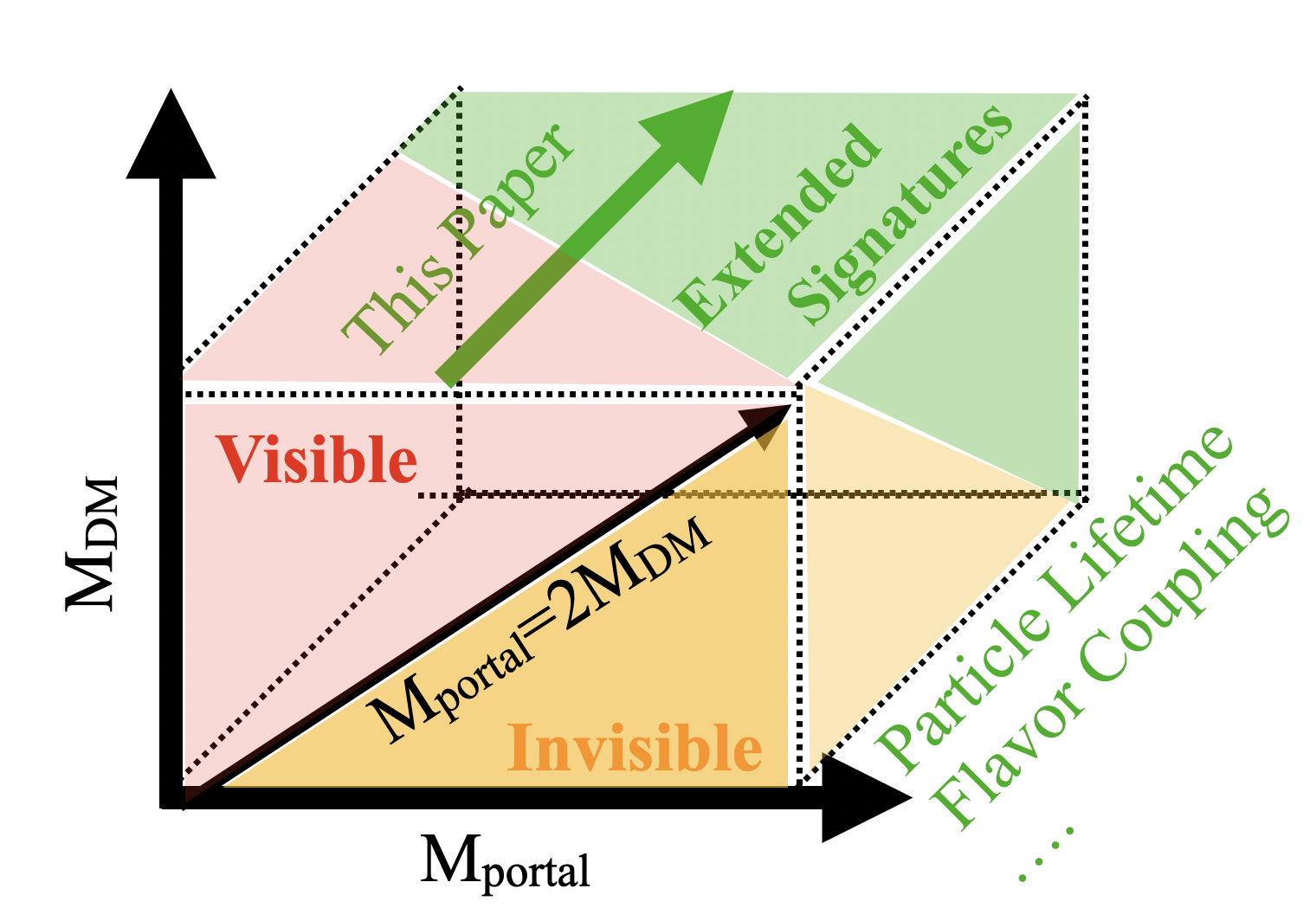}
    \caption{Visualization of the motivation for this white-paper. Typical dark sector searches focus on invisible (orange) and visible (red) final states, whereas in this white-paper we highlight the extended dark sector searches obtained by adding a variety of additional well motivated features  indicated by 3rd green axis. }
    \label{fig:diagram}
\end{figure}

Figure~\ref{fig:diagram} illustrates the vision for this paper. With a more complex dark sector, a variety of different final states are produced. Many of these final states go beyond the conventional signatures leading to combinations of invisible and visible final states. Furthermore, an extended dark sector can also lead to final states, consisting of displaced dark matter decays, partially visible decays, and final states resulting from usual decays of mesons.

Since there are countless ways to make the dark sector more complicated, we focus on explanations that are well motivated simple extensions of the benchmark model.  In particular, we consider simple extensions of dark sector models, which lead to different final states from the benchmark models. These extensions act as a test to ensure that the planned experimental program will remain robust to unexpected final states. Additionally, we consider other extensions, or modifications, of the dark sector models that are motivated by recent anomalies observed within the data. Anomalous features within the data could be an indication of some fundamental unobserved effect that is indirectly pointing to dark matter.  Unlike the benchmark models that aim to explore parameters, which can lead to dark matter relic densities compatible with observations, this paper seeks to connect dark matter production with other phenomenological features. As a result, we will establish new benchmarks that can be both compatible with dark matter production and explain other fundamental physics questions. 

While this paper is not inclusive, we will highlight a number of recent, well motivated, anomalous features in the data. Additionally, we will also consider dark sector extended models for which there is strong theoretical motivation and which lead to different final states to the standard benchmarks. Despite not being inclusive, we aim to highlight a variety of the most intriguing final states that go beyond the simplest explanations of dark matter and further motivate searches for dark sector physics in the next decade.

\section{Motivations and Scope}
\label{sec:motivation}

The Standard Model has a relatively complex structure. The fermions come in three copies, with fermions in each generation belonging to several different representations of the $SU(3)\times SU(2)\times U(1)$ gauge group. Even the gauge group is far from minimal, being a product of three simple groups. It stands to reason that dark matter could also be part of a more complex dark sector structure. As we will show below, using several sample dark sector models, the existing and planned experimental program can be used to probe also non-minimal dark sectors with robust coverage and clear potential for discovery.  

The non-minimal dark sector structure can be motivated by experimental anomalies. For instance, the dark sector models discussed in Section \ref{sec:g-2} were built to explain the $(g-2)_\mu$ anomaly, and as such have a nontrivial flavor structure, coupling either only to muons or at the very least with a different strength to muons than to the first and third generation leptons. Sections \ref{sec:iDM} and \ref{sec:SIMPs} contains examples of non-minimal dark matter models, motivated either by reducing the present observational constraints, as in the case of inelastic DM, or by exploring nontrivial cosmological histories (as well as non-trivial experimental signatures), as in the case of the strongly interacting massive particles. Section \ref{sec:axions} on the other hand, focuses on non-minimal axion models, which for instance can have flavor violating couplings if the PQ symmetry is part of the global flavor symmetry of the SM. 

Two underlying questions typically underpin the decision to extend a dark sector model beyond a minimal benchmark. First, is there a reason for well-motivated theoretical extensions of the Standard Model? Second, can the dark sector be extended to resolve anomalous observations within data? Before embarking on a detailed analysis of benchmarks, we summarize the most compelling motivations currently being considered.

\subsection{Theoretical motivations}
Theoretical motivations for an extended sector are governed by trying to extend the dark sector to introduce additional, well motivated dark sector dynamics or by trying to introduce a new mechanism to help explain the observed dark matter relic density. 

In section~\ref{sec:iDM}, we introduce the inelastic dark matter as a typical extended model that motivates an increase in the variety of signatures beyond the simplest possible signatures. Here, we modify the dark sector to decay into an unstable particle that subsequently decays to the dark matter particle \cite{Tucker-Smith:2001myb,Izaguirre:2015zva,Izaguirre:2017bqb,Berlin:2018jbm}. The motivation for extending the dark sector with an additional unstable particle is twofold. First, it arises naturally in a broad range of models. Second, it gives rise to new signatures and evades existing bounds from other experiments, such as direct detection.  There are many other dark sector models that give rise to new signatures including dark sector models built on dark QCD, where there is the possibility of dark showers. 

A theoretical extension of dark sector models comprises the flavor violating axion portal. This portal gives rise to new rare kaon, $B$ and $D$ decays and can potentially be related to the QCD axion.  This model is discussed in sections~\ref{sec:axions}. Many other models exist, including charged mediators~\cite{Ghosh:2022zef}, flavor violating neutrino interactions~\cite{Arguelles:2022xxa}, extended dark sector Higgs boson couplings~\cite{Rizzo:2022qan}, dark QCD sector~\cite{Albouy:2022cin}, modified dark photon couplings~\cite{Dutta:2022qvn}, and rich dark sectors~\cite{Dienes:2022zbh}. 

An additional motivation for an extended dark sector originates from a modified scenario to produce the correct dark matter relic density. Beyond traditional thermalization mechanisms, dark matter production can be significantly altered through new processes that change the number of particles or impede the thermalization process. One such example is 3$\rightarrow$2 processes, where we can allow for different couplings between matter and dark matter and still yield a dark matter relic density comparable to what is observed~\cite{Hochberg:2014dra}.  These processes were one of the original motivations for SIMP models discussed in section~\ref{sec:SIMPs}. Another alternative scenario consistent with the current dark matter relic density is the freeze-in scenario, where dark matter and matter interactions are so weak the dark matter never attains thermal equilibrium~\cite{Hall:2009bx}.

\subsection{Anomalies}

In the past 10 years, a variety of physics anomalies have appeared in the data. For many of these anomalies, the explanations have also appeared that connect the anomalies to the physics of the dark sector. This section highlights several compelling physics anomalies that can be studied experimentally within the next 10 years. 

{\bf (g-2)$_\mu$:} In 2021, the g-2 experiment confirmed the observation of a deviation in the value of the anomalous magnetic moment of the muon, $(g-2)_\mu$, from its  expected value \cite{Muong-2:2006rrc}. This observation, combined with the results of the previous experiment, led to a deviation of more than 4 standard deviations from the Standard Model expected observation. The lack of a deviation in the electron anomalous magnetic moment motivates an explanation of new physics, which couples to muons and potentially to other heavy particles \cite{Dreyer:2021aqd,Apyan:2022tsd,Capdevilla:2021kcf}.  Dark sector models capable of explaining $(g-2)_\mu$ consist of spin-$0$ and spin-$1$ portals where the electron coupling is heavily suppressed or turned off. Two well motivated models are discussed in section~\ref{sec:g-2}.

{\bf Flavor Anomalies:}  Recent measurements from LHCb indicate an ever-increasing deviation in the ratio of $B$ decays to final states with electrons compared to $B$ decays with muon final states. In particular, the recent observation of the ratio of $B\to K^{*}e^{+}e^{-}$ to $B\to K^{*}\mu^{+}\mu^{-}$ shows a deviation of nearly four standard deviations from the Standard Model prediction. Furthermore, other observables, such as $B\to D^{(*)}\ell\nu$ semileptonic decays \cite{HFLAV:2019otj}, also indicate similar deviations in the ratio of electron to muon or $\tau$ final states. Lastly, kinematic deviations, such as that of $P_{5}^{\prime}$ in $B\to K^{*}\mu^{+}\mu^{-}$ decays also show a deviation from what is expected~\cite{LHCb:2021trn}. Reconciling these anomalies with dark sector models again requires an extended dark sector, where the portal mediator couples differently to heavy flavor particles with respect to light flavor.

{\bf Xenon 1T Excess:} In the low energy region of the Xenon 1T experiment, an excess of events beyond the expected background is observed. This observation indicates either an unexplained background or the possibility of a very light dark matter candidate that has modified interactions yielding a negligible kinematic recoil. One of the leading explanations for the Xenon 1T excess is inelastic dark matter discussed in section~\ref{sec:iDM}~\cite{Harigaya:2020ckz}.

{\bf MiniBoone/MicroBoone Excess:} Low Energy recoils observed in the MiniBoone experiment has long been a motivation for sterile neutrinos. The recent results from MicroBoone have ruled out the conventional sterile neutrino hypothesis~\cite{MicroBooNE:2021zai}, yet they have neither eliminated nor explained the MiniBoone excess. One possible explanation for the features in both MicroBoone and MiniBoone is that the measured deviations could be due to a set of sterile neutrinos with modified flavor couplings~\cite{Arguelles:2022xxa}.

{\bf Neutron Lifetime Anomaly:} Measurements of the neutron lifetime obtained in ``beam'' measurements compared to those measured at low energy in ``bottle'' events substantially deviate from one another.  Physics, which leads to a modified neutron lifetime in different scenarios, could result from new particles originating from a mirror neutron. Active work is underway to confirm this anomaly through neutron conversion measurements at Oak Ridge National Lab~\cite{Broussard:2021eyr}. 

{\bf KOTO Anomaly:} Recently, the KOTO experiment observed an excess of events originating from invisible kaon decays with a $\pi^{0}$ in the final state \cite{KOTO:2018dsc}. The rate of events observed in KOTO significantly exceeds the predicted amount of events  from $K^0\to\pi^0\nu\nu$ decays within the Standard Model. Currently, the NA62 experiment is measuring the isospin related $K^\pm\to\pi^{\pm}\nu\nu$ decays to confirm if they see unexpected deviations. Similarly, KOTO is improving their measurement of the $K^0\to\pi^0+$MET channel.   If this unexpected rate remains anomalous, this could indicate invisible kaon decays to dark matter. Section~\ref{sec:axions} discusses a variety of models that would motivate anomalous kaon decay, with rates of $K^+$ and $K^0$ decays related by isospin symmetry, while models that break this relation were reviewed in  \cite{Goudzovski:2022vbt}.



\section{Benchmark Models}
\label{sec:models}
In the following section, we will highlight 4 benchmark models that motivate the use of extended dark sectors. In particular, we will talk about models motivated by the $(g-2)_\mu$ anomaly, inelastic dark matter, Strongly interacting Massive Particles (SIMPs), and axions with non-minimal couplings. 

\subsection{The models explaining $(g-2)_\mu$}
\label{sec:g-2}

\begin{figure}
    \centering
    \includegraphics[width=0.45\textwidth]{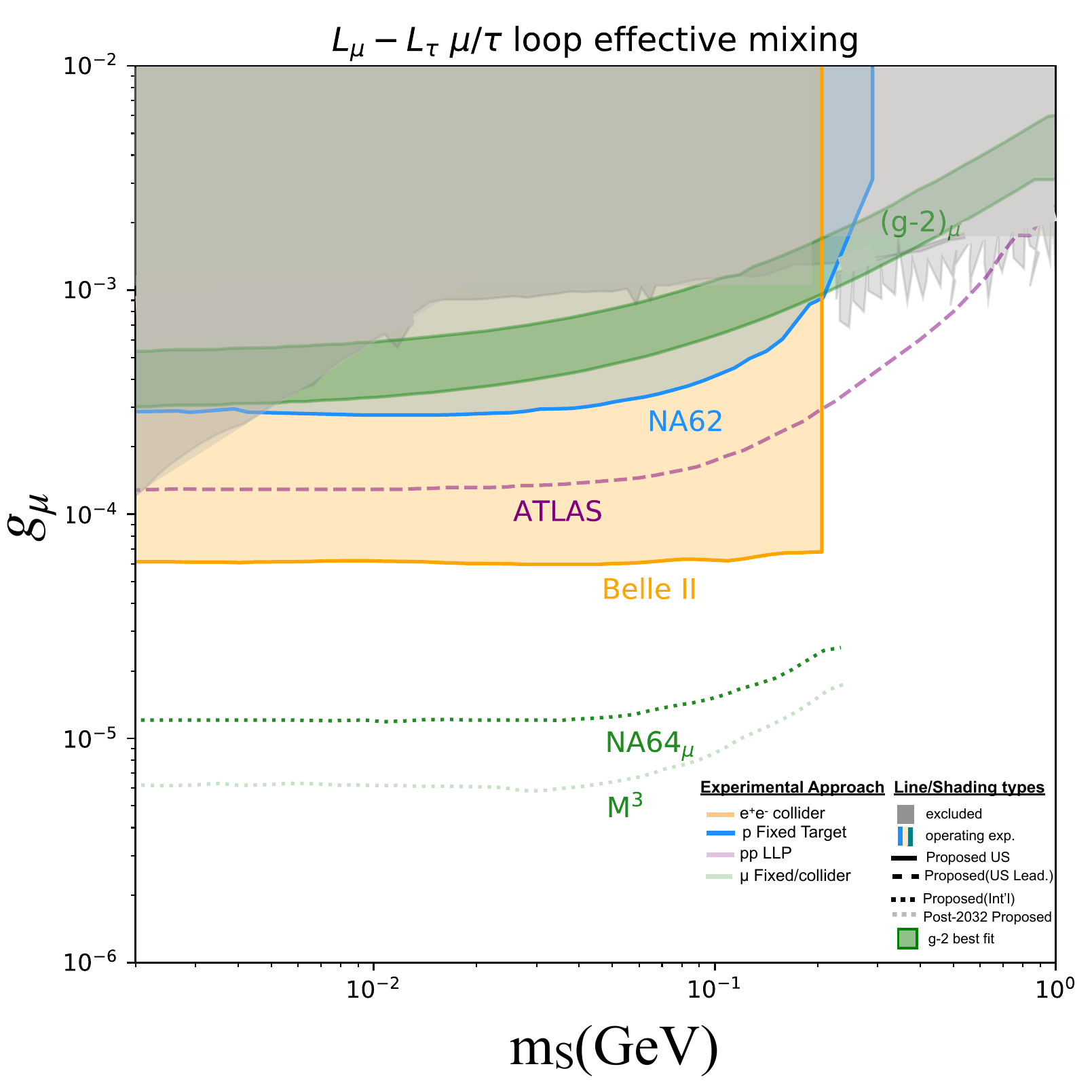}
    \includegraphics[width=0.45\textwidth]{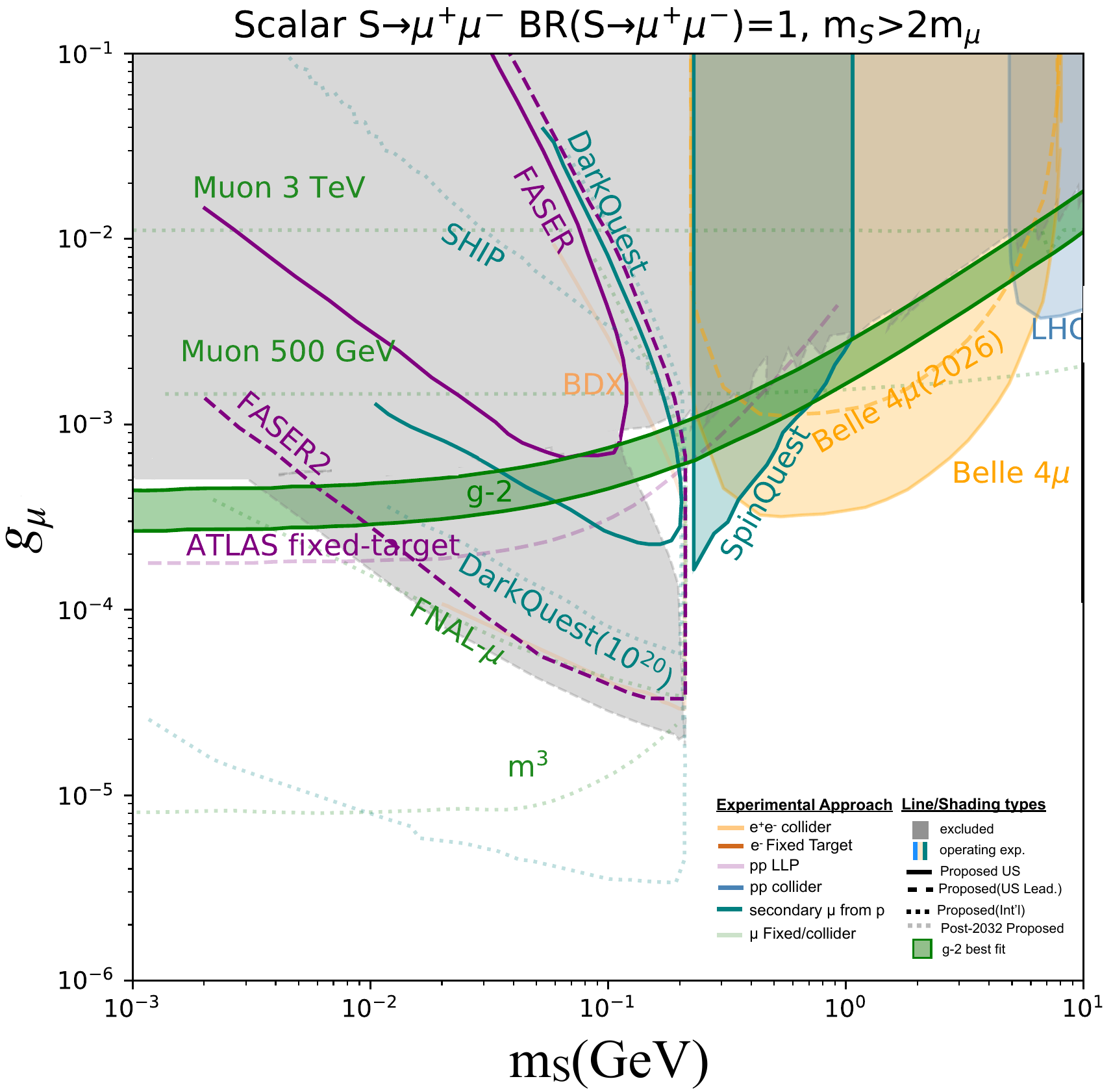}
    \caption{   \label{fig:Lmu:Ltau} Constraints (shaded regions) on the parameter space of the $L_\mu-L_\tau$ gauge boson (left) and light scalar $S$ coupling to muons (right), along with projected sensitivities for planned experiments. The parameter space that can explain the $(g-2)_\mu$ anomaly is shown as a green band. Shading implies the experiment is currently funded and under operation. The color, specified by the legend, indicates the type of experiment capable of probing the dark sector model. The line style indicates the level of US involvement, and the time scale of the planned experiment. 
    The projected sensitivities from planned or proposed experiments/measurements are shown with dotted (solid) lines in the left (right) panel. The figures are adapted from \cite{Greljo:2021npi,Capdevilla:2021kcf} with updated lines from\cite{NikitaTalk,GordonTalk}. }
    
\end{figure}


The experimental determination, $a^{\rm exp}_\mu$~\cite{Abi:2021gix}, of the anomalous magnetic moment of the muon, $a_\mu = \frac{1}{2}(g-2)_\mu$, differs by $4.2\sigma$ from  the consensus SM prediction, $a^{\rm SM}_\mu$~\cite{Aoyama:2020ynm}, obtained using dispersion relations, giving
\beq
\label{eq:Delta:amu}
\Delta a_\mu
=   a^{\rm exp}_\mu - a^{\rm SM}_\mu
=   \left( 251 \pm 59 \right) \times 10^{-11}.
\eeq
This long-standing discrepancy dates back to the first measurement performed by the Muon $g-2 $ collaboration at BNL~\cite{Muong-2:2006rrc}, and was recently strengthened by the measurements performed at  Fermilab~\cite{Abi:2021gix,Albahri:2021kmg,Albahri:2021ixb}. The experimental result is much closer  to the $a_\mu$ prediction that uses the BMW collaboration lattice QCD calculation of the hadronic vacuum polarization contribution~\cite{Borsanyi:2020mff}. In this case the SM prediction and experiment are consistent at $1.6\,\sigma$, however, the result should be confirmed by other lattice QCD groups (for recent lattice QCD results in the intermediate $q^2$ window, agreeing with the BMW collaboration's result in that range, and showing the discrepancy with a dispersive result, see \cite{Bhardwaj:2022qtk,Ce:2022kxy}).


If the discrepancy in $\Delta a_\mu$ \eqref{eq:Delta:amu} is assumed to be due to new physics, there are broadly two classes of new physics models that can lead to a contribution to $\Delta a_\mu$ of the required size (see, e.g., reviews in \cite{Lindner:2016bgg,Jegerlehner:2009ry,Crivellin:2018qmi,Athron:2021iuf,Altmannshofer:2022aml}). In all of the new physics models the new physics states contribute to $\Delta a_\mu$ through radiative corrections. The correct value of $\Delta a_\mu$ is obtained for qualitatively different parameter regimes depending on whether the chirality flip can occur on the internal new physics line.  The chirality flip on the internal line  requires at least two new physics fields, which can be quite massive, in the several TeV range, see Refs. \cite{Jana:2020joi,Chowdhury:2021tnm,Calibbi:2018rzv,Borah:2021khc,Acuna:2021rbg,Kowalska:2020zve,Chen:2020tfr,Kawamura:2020qxo,Calibbi:2019bay,Brdar:2021pla,Capdevilla:2021kcf,Capdevilla:2021rwo,Baum:2021qzx,Berger:2008cq,Stockinger:2006zn,Feng:2011aa,Cho:2011rk,Altmannshofer:2021hfu,Escribano:2021css,Arcadi:2021zdk,Bharadwaj:2021tgp,Dermisek:2021ajd,Baker:2021yli,Greljo:2021npi,Chen:2020jvl,Calibbi:2020emz,Belanger:2015nma,Arcadi:2021cwg,Perez:2021ddi,Heeck:2022znj,Arnan:2019uhr,Allanach:2015gkd,Crivellin:2021rbq,Endo:2020tkb,Fajfer:2021cxa,Harnik:2012pb,Dermisek:2013gta,Davoudiasl:2012ig} 
for concrete examples.

We focus on the second class of new physics models, in which the chirality flip occurs on the muon line, so that the contribution is suppressed by the muon mass. As a result the new physics states running in the loop are necessarily light. The constraints on two such models are shown in Fig. \ref{fig:Lmu:Ltau}. 

When building a new physics model, we envision a light mediator that can explain $\Delta a_\mu$ through a muon mediator coupling. In light of this, we can consider the thought process of trying to build a minimal model of a mediator that couples to muons and evades existing bounds. Given the large deviation in the observed $\Delta a_\mu$ measurement, a fairly large coupling to muons is required. This relatively large coupling implies that the models that couple directly to a broad range of particles, such as electrons or quarks, with couplings similar to that of the muon coupling, are already excluded. As a result, couplings of the mediator to other particles beside muon have to be either eliminated or heavily suppressed.

Table~\ref{tab:g2flow} illustrates this thought process, which drives the final state and models that can explain $\Delta a_\mu$. For a vector mediator, final states which include long-lived particles, photons, electrons, and hadrons, tend to be strongly constrained by the existing measurements. 
Models that in direct searches lead to such final states therefore have difficulty explaining $\Delta a_\mu$, while evading existing bounds. Furthermore, a mediator which is heavier than two times the muon mass $m_{S} > 2 m_{\mu}$ is already strongly constrained by the BABAR measurement.   Constraints for invisible final states are significantly weaker, leading to the possibility of a vector mediator, which couples to muons and a particle that decays invisibly, such as dark matter or, provided $m_{V} > 2 m_{\mu}$, neutrinos. 
As a result, the current most compelling vector model is the $L_{\mu}-L_{\tau}$ model, and the best experimental final state results from a mediator that couples to muons, and decays invisibly (for a more complete discussion of vector mediator models, including constraints from neutrino experiments, see \cite{Greljo:2022dwn}). Experiments, such as NA64${}_{\mu}$ \cite{Sieber:2021fue}, and M$^{3}$\cite{Kahn:2018cqs}, that are designed for missing momentum searches with muon beams are the most sensitive ones.

A scalar mediator that couples to muons has a larger variety of possible final states with weak couplings that are are not yet fully excluded. The scalar enables the possibility of loop suppressed final states yielding di-electron, and di-photon final states for a light mediator which only couples directly to muons, where the mediator is lighter that twice the muon mass, $m_{S} < 2 m_{\mu}$. For a heavier mediator, bounds on the mediator decaying to muons and pions do not yet exclude the $(g-2)_\mu$ best fit. Despite being able to decay directly to muons, the couplings are sufficiently small that mediator decays to muons will have a lifetime $c\tau\sim 1mm$ and have not yet been excluded. Fortunately, this coupling is sufficiently large to test with secondary muon beams. The future secondary muons from proton beam dumps, such as at the SpinQuest experiment at Fermilab, will be able to probe the critical unexplored parameter space in the near future. Finally, when considering scalar mediator models, it is also important to realize that a fully complete UV model may require new SM-charged final states above the weak scale that could potentially impact the $\Delta a_\mu$ measurement. 


\begin{table}
\begin{tabular}{ |p{2.1cm}|p{1.54cm}|p{1.5cm}|p{0.8cm}|p{1.54cm}|p{1.54cm}|p{1.58cm}|p{1.54cm}|  }
 \hline\hline
 & \multicolumn{3}{|c|}{Invisible} &  \multicolumn{4}{|c|}{Visible} \\
 \hline\hline
Final State/ Mediator & Long-Lived & Neutrinos ($\nu\nu$) & DM ($\chi\chi$) & Photons ($\gamma\gamma$) & Electrons ($e^{+}e^{-}$) & Muons ($\mu^{+}\mu^{-}$) & Hadrons ($\pi^{+}\pi^{-}$)  \\
 \hline
\multirow{2}{4em}{Vector} & No & Yes & Yes & No & No & Yes  & No  \\
				        &       &        &       &       &       & \footnotesize{$m_{V}>2m_{\mu}$} &  \\
\cline{2-8}
& \multicolumn{7}{|L{10cm}|}{
\begin{itemize}
\item $L_{\mu}-L_{\tau}$ gauge boson: UV complete, automatic coupling to neutrinos, easy to couple to DM. 
\item Challenging to build viable models with sizable couplings of vector mediator to electrons or hadrons (gauge anomalies, constraints from neutrino physics).
\end{itemize} 
} \\
\hline
\hline
\multirow{2}{4em}{Scalar} & Yes  						& Yes & Yes & Yes  			      & Yes 					 & Yes  & Yes   \\
					& \footnotesize{$m_{S} < 2m_{\mu}$} & 	 & 	   & \footnotesize{$m_{S} < 2m_{\mu}$} & \footnotesize{$m_{S} < 2m_{\mu}$} &  \footnotesize{$m_{S} > 2 m_{\mu}$} & \footnotesize{$m_{S} > 2 m_{\pi}$} \\
\cline{2-8}
& \multicolumn{7}{|L{10cm}|}{
\begin{itemize}
\item All minimal signatures can be realized in scalar simplified models.
\item  UV complete models require new SM-charged states above weak scale with special flavor structure (such states can in principle also affect $(g-2)_\mu$).
\item More phenomenological studies needed to chart the parameter space.
\end{itemize}
} \\
\hline
Signature &  \multicolumn{3}{|c|}{Missing Energy} &  \multicolumn{4}{|c|}{Prompt or Displaced Resonance} \\
  \hline\hline
\end{tabular}
\caption{A summary of experimental signatures in direct searches for a light vector or scalar mediator explanations of $(g-2)_\mu$ \label{tab:g2flow}\cite{BatellSlides}. }
\end{table}

Fig.~\ref{fig:Lmu:Ltau} shows current bounds on the two models discussed, above, i.e., the vector and scalar mediated models that can explain $\Delta a_\mu$. The  left panel shows constraints on the   gauged $L_\mu-L_\tau$ model in the parameter space of the $Z'$ (commonly denote also as $X$) gauge boson mass, $m_X$, and the gauge coupling $g_X$ \cite{He:1990pn,He:1991qd,Altmannshofer:2014cfa,Altmannshofer:2019zhy,Greljo:2021npi,Alonso-Alvarez:2021ktn,Cen:2021iwv,Coloma:2020gfv,Heeck:2018nzc,Biswas:2021dan}. The parameter region that leads to the explanation of $(g-2)_\mu$ is within solid green lines, with the dashed line denoting the central value. The constraints from other experiments (color shaded) regions are avoided for $Z'$ masses of around $100$ MeV. The next generations of experiments (dotted lines) are expected to explore fully the parameter space relevant for $(g-2)_\mu$. Note that there are other anomaly free gauged  $U(1)_X$ models that can explain the $(g-2)_\mu$ anomaly, however, $L_\mu-L_\tau$ is the least constrained  \cite{He:1990pn,He:1991qd,Altmannshofer:2014cfa,Altmannshofer:2019zhy,Greljo:2021npi,Alonso-Alvarez:2021ktn,Cen:2021iwv,Coloma:2020gfv,Heeck:2018nzc,Biswas:2021dan,Greljo:2022dwn}.

The right panel in Fig.  \ref{fig:Lmu:Ltau} shows, in grey, the present constraints on a scalar singlet $S$  of mass $m_S$, coupling to muons \cite{Capdevilla:2021kcf}
\beq
\label{eq:Smuon}
{\cal L}_{\rm int}=g_S S\bar \mu \mu_L+{\rm h.c.},
\eeq
while the $(g-2)_\mu$ preferred region is shown in green. This parameter range can be explored fully by the planned or proposed experiments (shaded areas, and solid lines).
In deriving the constraints and experimental reach in the right panel of Fig.  \ref{fig:Lmu:Ltau} only the minimal coupling in \eqref{eq:Smuon} is assumed, minimizing the experimental constraints. However, the interaction in \eqref{eq:Smuon} is not gauge invariant and needs to be UV completed. At low energies it would arise from dimension 5 operator $H \bar \mu_R L_2 S/\Lambda$ after Higgs obtains a vev. Here $\Lambda$ is the mass scale of the heavy states that are integrated out. In UV completions of \eqref{eq:Smuon} there are thus further constraints on the models from electroweak precision observables, and other measurements. More examples of such light new physics models that can explain $(g-2)_\mu$ include flavor violating ALPs \cite{Bauer:2019gfk}, ALPs with large couplings to photons \cite{Davoudiasl:2018fbb,Marciano:2016yhf,Buen-Abad:2021fwq}, contributions due to both photonic and leptonic couplings of ALPs \cite{Buttazzo:2020vfs}, or from ALPs coupling to a dark photon \cite{Ge:2021cjz}.

Quite generally, light new physics models explaining the $(g-2)_\mu$ anomaly are required to have a rather nontrivial flavor structure. As already mentioned, to avoid the stringent constraints from searches for light new physics, the couplings to electrons and first generation quarks need to be suppressed. Furthermore, any flavor violating couplings need to be highly suppressed, due to severe constraints from bounds on  lepton-flavor-violating transitions, such as $\mu\to e\gamma$ and $\mu \to 3 e$. For a generic flavor structure, lepton-flavor-violating transitions give bounds on the new physics scale that are much higher than what is required for $(g-2)_\mu$. Phrasing these constraints in terms of the effective new physics suppression scale for the dimension 5 dipole moment operators, $\mathcal{L}_{\rm eff} \supset - { e \,v}\, \bar \ell^i_{LL} \sigma^{\mu \nu} \ell^j_{RR } F_{\mu \nu}/{(4 \pi \Lambda_{ij})^2} + {\rm h.c.}$, where $v  ={246}${GeV} is the electroweak vev, and $i,j$ generational indices,  the new physics scale required to explain the $(g-2)_\mu$ anomaly  is $\Lambda_{22} \simeq {15}$\,{TeV}, while the absence of $\mu \to e \gamma$ implies $\Lambda_{12 (21)}  \gtrsim {3600}$\,{TeV}~\cite{MEG:2016leq}. The flavor violating transitions therefore need to be significantly suppressed, either by ad-hoc flavor alignment of new physics couplings, or through the use of symmetries, such as the $U(1)_X$ gauge symmetry in the case of light $Z'$ explanations for $(g-2)_\mu$ \cite{Greljo:2021npi}.

In summary, given the strong bounds from existing experiments, the least constrained models tend to uniquely couple directly to muons.  As a result, a muon beam either through secondary muons in a proton beam, or directly produced is critical to definitively probing the dark sector parameter space motivated by the $(g-2)_\mu$ anomaly. There are a variety of experiments built on such principles that are being considered, which will likely have the ability to put significant constraints on the entire remaining parameter space that can explain the $(g-2)_\mu$ anomaly.


\subsection{Inelastic DM}
\label{sec:iDM}

When considering theoretical extensions of dark sector models, a simple modification of minimal dark matter models arises from the possibility that the dark sector consists of several states. The heavier states are unstable and can be searched for using their decay products. Furthermore, if the dark sector particles are lighter than half the mediator and the dark matter interaction strength, $\alpha_{D}$, is large, the dark sector particles are readily produced both in the laboratory and in the early universe. 

A particularly interesting example is the inelastic dark matter, in which the mediator between the dark sector and the SM couples inelastically to the two DM states, $\chi_{1,2}$ \cite{Tucker-Smith:2001myb,Izaguirre:2015zva,Izaguirre:2017bqb,Berlin:2018jbm}. This can occur for a dark photon mediator, $A^{\prime}$, yielding the interaction to the two dark states given by
\begin{equation}
\label{eq:L:inelastic}
\mathcal{L} \supset i e_{D} A^{\prime}_{\mu} \bar{\chi}_{1}\gamma^{\mu}\chi_{2},
\end{equation}
while the diagonal couplings of $A'$ to just $\chi_1$ or just $\chi_2$ vector currents are either absent or highly suppressed. Often the two dark matter mass eigenstates, $\chi_{1,2}$, are almost degenerate, but with mass splitting that is large enough to have phenomenological consequences. Inelastic dark matter was first introduced to explain the DAMA dark matter excess \cite{Tucker-Smith:2001myb}. Choosing judicially the mass splitting of ${\mathcal O}(100\,\text{keV})$  such that dark matter scattering in direct detection experiments was kinematically allowed for heavy nuclei and forbidden for light nuclei, this then evaded the most stringent direct detection bounds at the time.  Inelastic dark matter is also a simple light thermal DM model that does not lead to late time annihilations and thus evades the otherwise very stringent bound on light DM scenarios due to distortions of cosmic microwave background (CMB) \cite{Planck:2015fie,An:2016kie,Slatyer:2015jla}.

A simple realization of the inelastic DM is a model with two Weyl fermions, $\eta$ and $\xi$, oppositely charged under the broken $U(1)_{D}$ symmetry. The mass terms in the Lagrangian are then given by, 
\begin{equation}
-\mathcal{L} \supset m_{D} \eta\xi + \frac{\delta_{\eta}}{2}\eta^{2}+\frac{\delta_{\xi}}{2}\xi^{2}, 
\end{equation}
where we assume the hierarchy $m_{D} \gg \delta_{\eta,\xi}$. The Dirac mass term is allowed by the $U(1)_D$ symmetry, while $\delta_{\eta,\xi}$ give the symmetry breaking (this can be either explicit, if $U(1)_D$ is global, or due to a vacuum expectation of a scalar field carrying double the charge of $\eta, \xi$ under $U(1)_D$). In the limit of exact $U(1)_D$ the two Weyl fermions combine into a single Dirac fermion, while for  $\delta_{\eta,\xi}\ne0$ we have a pseudo-Dirac fermion, i.e., two almost degenerate Majorana fermions.  The two Majorana fermion mass eigenstates are given by,
\begin{equation}
\chi_{1} \simeq i\left(\eta-\xi\right)/\sqrt{2}, \qquad
\chi_{2} \simeq  \left(\eta+\xi\right)/\sqrt{2},
\end{equation}
and have masses $m_{1,2} \simeq m_D \mp \frac{1}{2}\left(\delta_{\eta}+\delta_{\xi}\right)$. 
The lightest of the two,  $\chi_{1}$, is stable and is assumed to be the DM. The heavier state, $\chi_{2}$, is unstable and eventually decays to $\chi_{1}$. The  $\chi_{2}\to \chi_{1}$ decay is characterized by the mass splitting
\begin{equation}
\Delta \equiv \frac{m_{2}-m_{1}}{m_{D}}\simeq\frac{\delta_{\eta}+\delta_{\xi}}{m_{D}},
\end{equation}
and leads to longer decay times, the smaller the splitting. For gauged $U(1)_D$ the decay is mediated by the corresponding gauge boson, which couples inelastically to the two mass eigenstates, cf. Eq.  \eqref{eq:L:inelastic}. For heavy enough dark photon both $\chi_{2}$ and $\chi_{1}$ can be produced from dark photon decays. 
Furthermore, $\chi_{2}$ can decay to $\chi_{1}$ and standard model particles  via a tree level exchange of an off-shell dark photon,  $\chi_{2}\rightarrow\chi_{1}A^{\prime*}\rightarrow\chi_{1}f\bar{f}$. The partial decay width for decays to lepton pairs, for instance, is given by 
\begin{equation}
\Gamma\left(\chi_{2}\rightarrow\chi_{1}\ell^{+}\ell^{-}\right)\simeq \frac{4 \epsilon^{2} }{15\pi } \alpha_{\rm em} \alpha_{D} \frac{m_{1}^5}{m^{4}_{A^{\prime}}} \Delta^{5}.
\end{equation}


\begin{figure}[t]
\centering
\includegraphics[width=15cm]{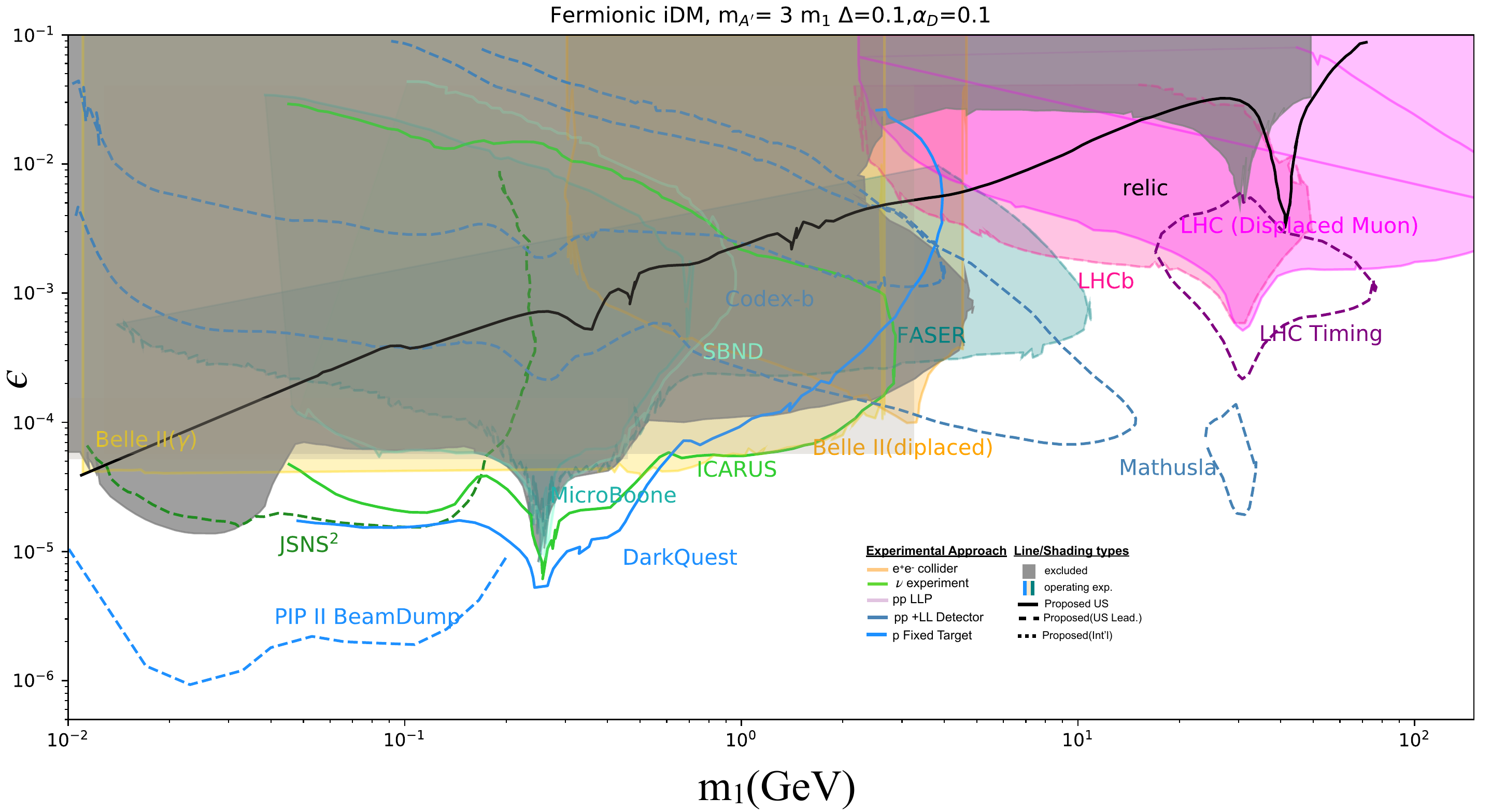}
\caption{\small Constraints on inelastic DM (shaded gray)  as functions of DM mass $m_1$ and the dark photon mixing parameter,  $\epsilon$, taking dark photon mass to be $m_{A'} / m_1$ = 3, the iDM mass splitting $\Delta = 0.1$, and $\alpha_D = 0.1$. The projected reach of low-energy accelerators, future displced searches,and  neutrino exepriments, amongs other,  are show with color coded final states and linestyled indicative of their usa status~\cite{Belle-II:2018jsg},~\cite{Berlin:2018pwi}, ~\cite{Izaguirre:2015zva},~\cite{Izaguirre:2017bqb}, JSNS$^2$ (red)~\cite{Jordan:2018gcd}, ~\cite{Izaguirre:2017bqb}, ~\cite{Berlin:2018pwi,Berlin:2018bsc,LDMX:2018cma}, and future sensitivity from electroweak precision tests (EWPT) at the LHC (dark teal)~\cite{Curtin:2014cca}. Dashed lines show the projected reach at LHC based experiments with US leadership, CODEX-b, MATHUSLA, FASER, as well as HL-LHC searches at ATLAS and CMS (see \cite{Berlin:2018jbm} for details). Black solid line denotes for which parameters $\chi_1$ is a thermal relic that has the right DM abundance. 
Figure adapted from \cite{Berlin:2018jbm} with updates from \cite{Krnjaic:2022ozp}. \label{fig:iDM}}
\end{figure}



For typical values of $\Delta=0.1$, $\epsilon=10^{-4}$, $\alpha_{D}=0.1$, $m_{A^{\prime}}=100$~MeV and $\chi_2$ mass comparable to $m_{A'}$, the $\chi_2$ lifetime is roughly $c\tau_{\chi_2}\sim 1000$~km.  For a light $\chi_{2}$, or a small mass splitting the decays to fermions are often limited to the electrons due to kinematic constraints. Despite the additional unstable particle, the DM relic density calculation is not significantly impacted by $\chi_{2}$ when compared to the case of Dirac fermion DM coupling to a dark photon. In the iDM scenario thermal freeze-out occurs slightly earlier, and at larger couplings, due to the fact that $\chi_{2}$  is heavier, and its abundance
depletes faster than for $\chi_{1}$, with co-annihilation of $\chi_{1}$ and $\chi_{2}$ dominating the DM thermalization in the early universe. Consequently, larger mass splittings require larger dark matter interaction strengths in order to satisfy relic density constraints. 

Inelastic dark matter is a compelling extension of the benchmark dark sector models which results in diminished impact of some of the dark matter searches. In particular, the cross section for dark matter direct detection is heavily suppressed in inelastic dark matter models \cite{Bramante:2016rdh}. Furthermore, inelastic DM provides a mechanism to suppress annihilations at late times in order to alleviate strong bounds from measurements of the
CMB. As a result of these restrictions, collider and beam dump experiments provide the strongest constraints on the possibility of inelastic dark matter. 

Figure~\ref{fig:iDM} shows the sensitivity for inelastic dark matter using a benchmark mass splitting of $\Delta=0.1$, and dark photon gauge coupling strength $\alpha_{D}=0.1$ \cite{Berlin:2018jbm}. A wide variety of experiments are capable of covering the phase space available, with beam dump experiments dominating the lower region of parameter space, and experiments at the Large Hadron Collider dominating the parameter space beyond the dark matter mass of a few GeV (see also \cite{Duerr:2019dmv,Duerr:2020muu}). An important observation is that the current set of current and planned experiments will be capable of completely covering the allowed parameter space that can explain the dark matter relic density.


\subsection{SIMPs}
\label{sec:SIMPs}
Another theoretical extension of the dark sector is to envision extended, standard model-like properties, characteristic with the rich particle structure present in the standard model. For example, the dark sector could contain dark quarks charged under a confining group, for instance, the $SU(N_c)$ gauge group  \cite{Berlin:2018tvf} (for a review of other possibilities see, e.g., \cite{Kribs:2016cew}). Similarly to QCD the spectrum of dark sector particles is then composed of strong interaction resonances,  with pions $\pi_D$ the lightest, followed by vector resonances $V_D$, and heavier states. These are the analogues of $\pi$, and $\rho, \omega, \phi$ resonances in the QCD.

Correct DM relic abundance of neutral $\pi_D$, made stable, e.g., by $G$-parity in the dark sector, requires that the dark sector couples to the SM through a portal such as the dark photon. Without such couplings to the SM the number changing processes such as $2\pi_D \to3\pi_D$ would heat up the dark sector leading to phenomenologically non-viable cosmology \cite{Carlson:1992fn,Hochberg:2014dra,Hochberg:2014kqa}. In addition to resolving cosmological concerns, the presence of the dark photon is essential to enabling the production of SIMP particles in a laboratory environment by enabling interactions of dark sector particles to standard model particles. Moreover, the existence of the dark photons and SIMPs naturally lead to a weak Standard model coupling that yields Long Lived Particles. 

\begin{figure}[t]
\centering
\includegraphics[width=9cm]{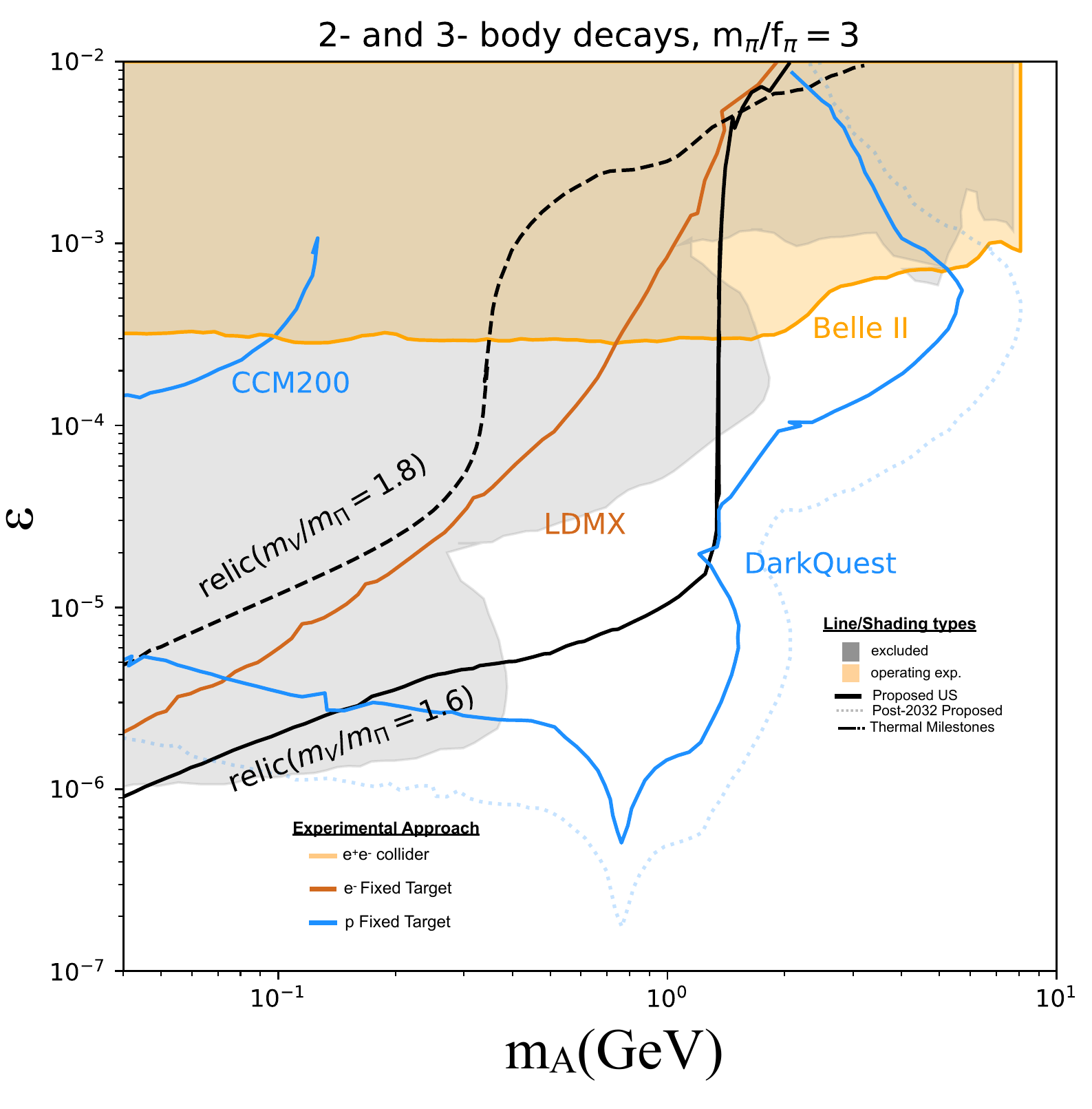}
\caption{\small An example of exclusions in the SIMP parameter space for a QCD-like strongly coupled dark sector gauged under a dark photon U(1) that induces interactions to the SM through the mixing parameter $\epsilon$.  The ratio of dark pion mass to decay constant is fixed to $m_\pi/f_\pi$, as well as the dark U(1) fine structure constant, $\alpha_D = 10^{-2}$, the ratio of dark photon and pion masses,  $m_{A'} / m_\pi =
  3$, and vector to pion masses, $m_V/m_\pi = 1.8$, while the dark photon mass $m_{A'}$ is varied. All hidden sector vector mesons are assumed to decay to the SM particles via two-body $V\to \ell^+\ell^+$ or three body $V\to \pi \ell^+\ell^-$ transitions. 
The grey shaded regions are
  excluded by BaBar~\cite{BaBar:2017tiz}, E137~\cite{Bjorken:1988as},
  Orsay~\cite{Davier:1989wz}, and searches for DM scattering at
  LSND~\cite{deNiverville:2011it}, E137~\cite{Batell:2014mga}, and
MiniBooNE~\cite{MiniBooNE:2017nqe}. The solid lines give the projected reach of Belle II~\cite{Belle-II:2018jsg}
  (orange),  DarkQuest~\cite{Apyan:2022tsd}
  (blue), and LDMX~\cite{Izaguirre:2014bca} (brown).  Dark pions constitute all of the observed DM abundance on solid (dashed) black
  contours for $m_V/m_\pi = 1.8$ ($1.6$), while DM is overabundant
  below these lines. 
  Adapted from \cite{Berlin:2018tvf,Apyan:2022tsd}. 
\label{fig:SIMP}}
\end{figure}

Because of the interactions with the SM the SIMPs can be searched for in lab experiments \cite{Hochberg:2015vrg,Berlin:2018tvf}. Fig. \ref{fig:SIMP} shows the present constraints and projected reach of experiments for an example of QCD-like SIMPs, that interacts with the SM trough a dark photon mediator. Further examples and more thorough discussion can be found, e.g.,  in Refs. \cite{Berlin:2018tvf,Bernreuther:2022jlj,Albouy:2022cin,Arrington:2022pon,Apyan:2022tsd,Baltzell:2022rpd,Asadi:2022njl,Cheng:2021kjg,Bernreuther:2019pfb,Kribs:2018oad,Kribs:2018ilo}. 

\subsection{Axions with non-minimal couplings}
\label{sec:axions}

Pseudo Nambu-Goldstone bosons (pNGBs) have masses that are naturally much smaller than the scale of new physics,  and can  act as an invaluable window to the high scale dynamics. The pNGBs or  axion-like particles (ALPs) arise whenever an approximate global symmetry is spontaneously broken. A well motivated example of an ALP is the QCD axion: it simultaneously solves the strong CP problem~\cite{Peccei:1977hh,Wilczek:1977pj,Weinberg:1977ma}, and is a valid cold dark matter candidate~\cite{Preskill:1982cy,Abbott:1980zj,Dine:1982ah}. Other theoretically motivated scenarios with light ALPs include low scale supersymmetry, composite Higgs models, models of dark matter freeze-out, and models of electroweak baryogenesis (see, e.g., Refs.~\cite{Kilic:2009mi,Bellazzini:2012vz,Ferretti:2013kya,CidVidal:2018blh,Jeong:2018ucz}).

At low energies the effective Lagrangian of ALP interactions with the SM is given by
\begin{equation}
\mathcal{L}_{\text{ALP}}=\frac{(\partial a)^2}{2}-\frac{m_a^2a^2}{2}+\mathcal{L}_{\text{ALP-gauge}}+\mathcal{L}_{\text{ALP-}f} \label{L_ALP}\, .
\end{equation}
The  couplings of $a$ to the SM fields arise from integrating out heavy fields at scale $\Lambda_{\text{UV}}\sim 4\pi f_a$ resulting in dimension 5 interactions,
\begin{align}
\label{eq:alpgauge}
   \mathcal{L}_{\text{ALP-gauge}}&=\frac{N_3\alpha_s}{8\pi f_a} a G_{\mu\nu}^a\tilde{G}^{a\mu\nu}+\frac{N_2\alpha_2}{8\pi f_a} a W_{\mu\nu}^i\tilde{W}^{i\mu\nu}+\frac{N_1\alpha_1}{8\pi f_a} a B_{\mu\nu}\tilde{B}^{\mu\nu}\,,
   \\
   \label{eq:ALP-f}
   \mathcal{L}_{\text{ALP-}f} &= 
\frac{\partial_\mu a}{2 f_a} \, \bar f_i \gamma^\mu \big( C^V_{f_i f_j} + C^A_{f_i f_j} \gamma_5 \big) f_j. 
\end{align}
Here $G^a$ denote SM gluons, $W^i$ the $SU(2)_L$ gauge bosons, $B_\mu$ the hypercharge boson, and $f=u,d,\ell$ the SM fermions,  with gauge multiplets already decomposed in terms of the mass eigenstates, and $i,j=1,2,3$ the generational indices. The electroweak gauge invariance then requires $C^V_{d_i d_j} - C^A_{d_i d_j} = V^*_{u_k d_i } (C^V_{u_k u_l} - C^A_{u_k u_l}) V_{u_l d_j}$, where $V$ is the CKM matrix. The form of the ALP couplings to fermions, $C^{V,A}_{f_i f_j}$, is model dependent. While for simplicity in many cases these couplings are assumed to be flavor diagonal, this is not required in general, and flavor off-diagonal couplings can lead to enhanced sensitivity to ALP models.   In total the ALP interactions with the charged SM fermions and gauge bosons are specified by 33  parameters. 

%
%
%
%

\paragraph{Flavor violating QCD axion.} If PQ symmetry does not commute with the SM flavor group this will lead to flavor off-diagonal couplings of ALP with the SM fermions, see, e.g.,  Refs.~\cite{Bardeen:1977bd,Davidson:1981zd,Reiss:1982sq,Davidson:1983fy, Davidson:1984ik,Berezhiani:1990wn,Berezhiani:1990jj,Babu:1992cu,Feng:1997tn,Albrecht:2010xh,Ahn:2014gva,Celis:2014iua}. This scenario is particularly well motivated if PQ symmetry is involved in generating the hierarchy among the SM fermion masses. A minimal realization of such a scenario is the {\em axiflavon}~\cite{Ema:2016ops,Calibbi:2016hwq}, where the PQ symmetry is identified with the Froggatt-Nielsen $U(1)$ global symmetry responsible for the pattern of masses and mixings observed in the quark and charge lepton sectors. In this case the couplings of QCD axion are predicted to have both diagonal and off-diagonal couplings,  $C_{f_if_j}^V\propto \sqrt{m_{f_i} m_{f_j}}$, generating the flavor violating decays such as $K\to \pi a$, with $a$ effectively massless at colider energies and escaping the detector. In general, if the QCD axion has anarchic flavor structure, with all couplings $C_{f_i f_j}^{V,A}\sim {\mathcal O}(1)$, the most stringent collider constraints will be due to the $K\to \pi a$ decays, probing axion decay constants $f_a$ of the order of $10^{12}$ GeV, well above the astrophysics constraints. For other flavor patterns, other rare decays such as $B\to Ka, D\to \pi a$, etc., may become important. If UV theory is flavor universal the radiative corrections from ALP couplings to the up-quarks~\cite{Hall:1981bc,Freytsis:2009ct,Bauer:2020jbp} or to the $W$-boson~\cite{Izaguirre:2016dfi} will still introduce flavor violating couplings. However, these are suppressed enough that only much lower scales, at the level of $f_a\sim $TeV are probed via rare processes~\cite{Bauer:2021wjo,Bauer:2021mvw,Goudzovski:2022vbt}.

\begin{figure}[t]
\centering
\includegraphics[width=11cm]{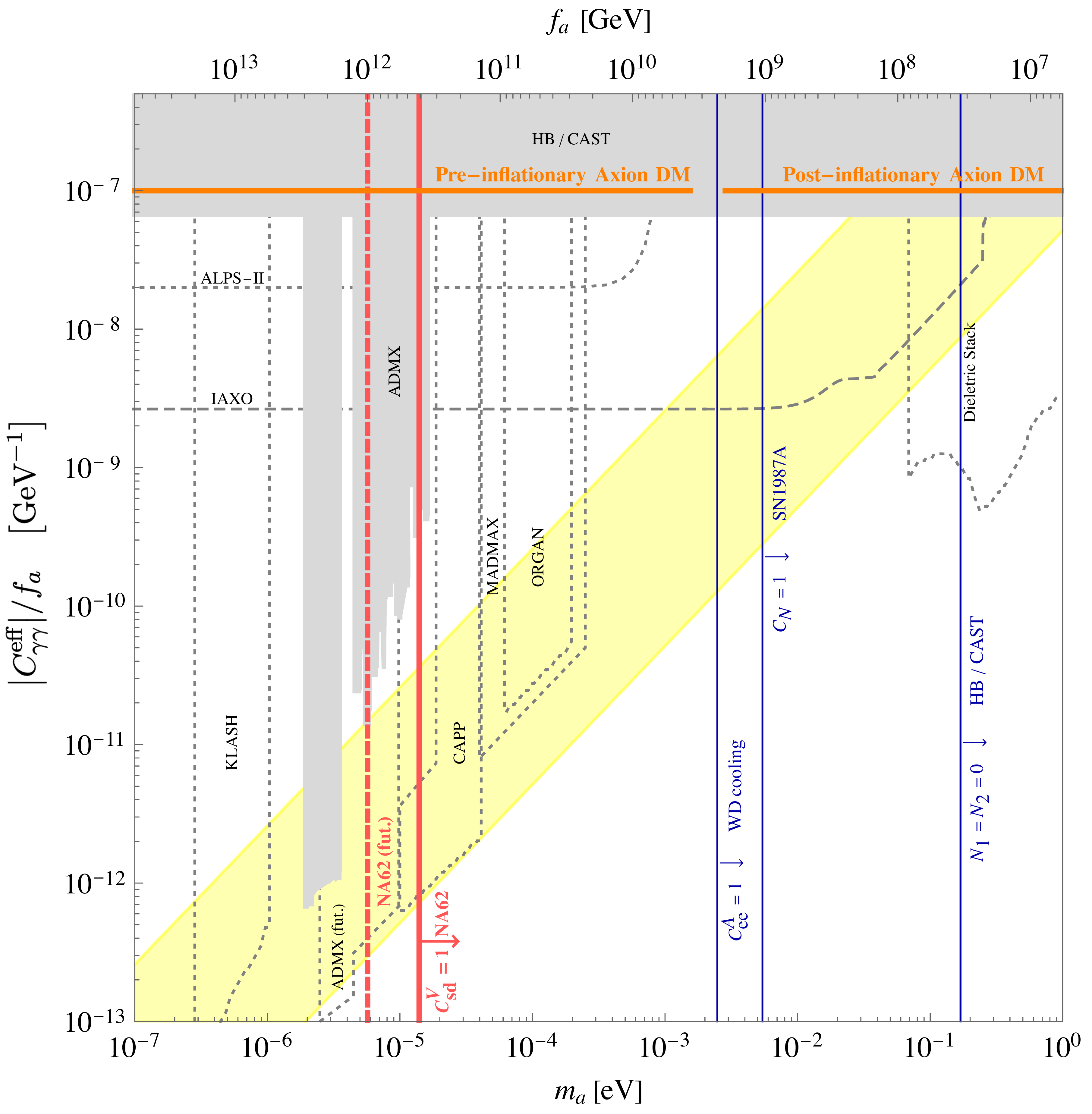}
\caption{\small Constraints on flavor violating QCD axion  in the plane of axion mass $m_a$ vs. the effective coupling to photons $C_{ \gamma \gamma}^{\rm eff}$. Gray shaded regions are excluded from existing (dotted lines for planned) halo- and helioscopes and from HB star cooling. The standard QCD axion band is shown in yellow. To guide the eye the blue vertical lines show several constraints on $m_a/f_a$ for different choice couplings set to one; the HB/CAST bounds~\cite{CAST:2017uph, Ayala:2014pea} for $N_1 = N_2 = 0$ (i.e. the ``hadronic axion'' which couples to photons through its mixing with $\pi^0$), the bounds from star cooling via nucleon bremsstrahlung in SN1987A for $C_N =1$ \cite{Carenza:2019pxu}, and via electron bremsstrahlung in white dwarfs (WD) for $C^A_{ee} =1$ \cite{MillerBertolami:2014rka}. Red solid (dashed) lines show the present (future) constraints from NA62 for $C^V_{sd} =1$. Orange lines indicate the values of $m_a$ (and $f_a$) for which the QCD axion can fully account for the observed dark matter abundance, depending on whether the PQ  symmetry is broken before or after inflation (taken from Ref.~\cite{Gorghetto:2020qws}). Figure adapted from \cite{Goudzovski:2022vbt}. \label{fig:axionplot}}
\end{figure}

Fig. \ref{fig:axionplot} shows the constraints for an example of a flavor violating QCD axion for which we assumed a completely anarchic flavor structure for the couplings to the SM fermions, setting all the coefficients in \eqref{eq:ALP-f} to $C_{f_if_j}^{V,A}=1$, while the coupling to gluons was set to $N_3=1$. The solid (dashed) red vertical line shows the present (future) sensitivity from the most stringent rare meson decay constraint due to NA62 search for $K\to \pi a$.  We observe that this search already explores part of the viable parameter region where the flavor violating QCD axion is a cold dark matter candidate (orange horizontal lines). The mass of the QCD axion is given entirely due to the QCD anomaly and follows the usual QCD axion relation \cite{Gorghetto:2018ocs}
\beq
m_a =5.691(51)\mu \text{eV} \biggr(\frac{10^{12}\, \text{GeV}}{f_a}\biggr).
\eeq
In order to compare the sensitivity of searches for the $K\to \pi a$ decays with the QCD axion searches relying on the couplings to photons Fig. \ref{fig:axionplot} shows the constraints in the $ C_{\gamma\gamma}^\text{eff}/f_a$ vs. $m_a$ plane. 

In Fig. \ref{fig:axionplot} the effective coupling of the ALP to photons
\beq
   \mathcal{L}_{\text{ALP-photon}}=\frac{ \alpha}{8\pi f_a} C_{\gamma\gamma}^{\rm eff} a F_{\mu\nu}\tilde{F}^{\mu\nu},
\eeq 
 given by \cite{Bauer:2017ris, Bauer:2020jbp,Goudzovski:2022vbt}
\beq
\begin{split}
\label{eq:cgamgameff}
   C_{\gamma\gamma}^\text{eff}(m_a)
   &= N_1+N_2 +\sum_{q}\, 6\, Q_q^2\,C_{qq}^A(\mu_0)\,B_1(\tau_q)
    + 2\sum_{\ell} C_{\ell\ell}^A\,B_1(\tau_\ell)
    \\
   &\quad\mbox{}- (1.92\pm 0.04)\,N_{3} - \frac{m_a^2}{(m_\pi^2-m_a^2)} 
    \left[ N_{3}\,\frac{m_d-m_u}{m_d+m_u} + C_{uu}^A-C_{dd}^A \right] \,.
    \end{split}
\eeq
is treated as a free parameter, allowing also for values $C_{\gamma\gamma}^\text{eff}\gg 1$. Above, the loop function $B_1(\tau_f)$ depends on $\tau_f\equiv 4m_f^2/m_a^2$, and is $B_1\simeq 1$ for light fermions ($m_f\ll m_a$) while it decouples as $B_1\simeq -m_a^2/(12 m_f^2)$ for heavy fermions ($m_f\gg m_a$). The explicit form of $B_1(\tau)$ can be found in Ref.~\cite{Bauer:2020jbp}.
We neglected the threshold corrections coming from $W$-loops since these are negligible for $m_a\ll m_W$. The second line in Eq.~(\ref{eq:cgamgameff}) encodes the leading order chiral perturbation theory (ChPT) contribution of $N_3$ to the ALP di-photon coupling. Similarly, the ALP coupling to electrons is generated at one-loop from the ALP couplings to gauge bosons~\cite{Bauer:2017ris}.  A common notation in axion literature is also $g_{\gamma\gamma}=C_{\gamma\gamma}^\text{eff} \alpha /(2\pi f_a)$.

\begin{figure}[t]
\centering
\includegraphics[width=3in]{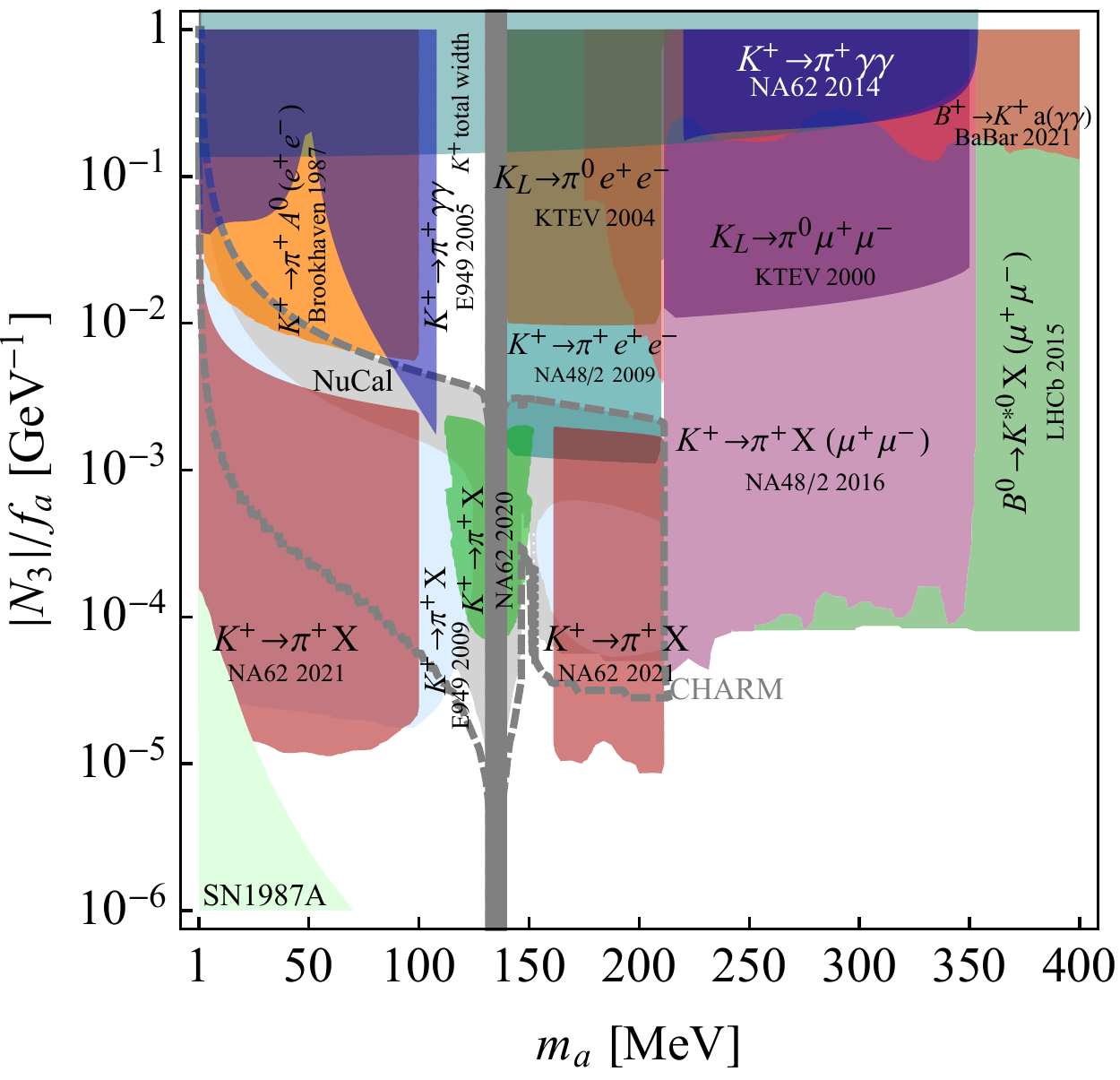}~~~~
\includegraphics[width=3in]{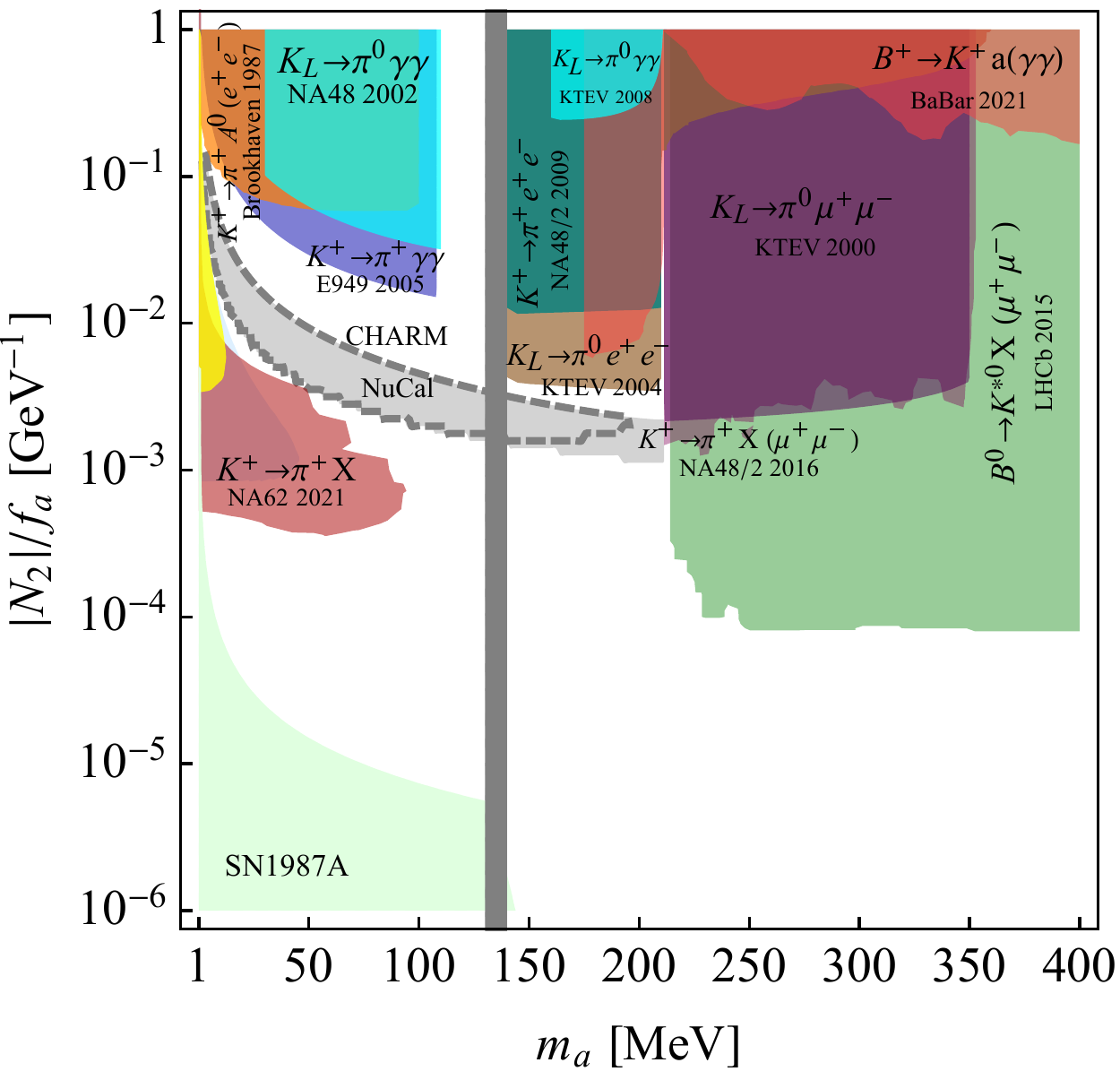}
\\
\includegraphics[width=3in]{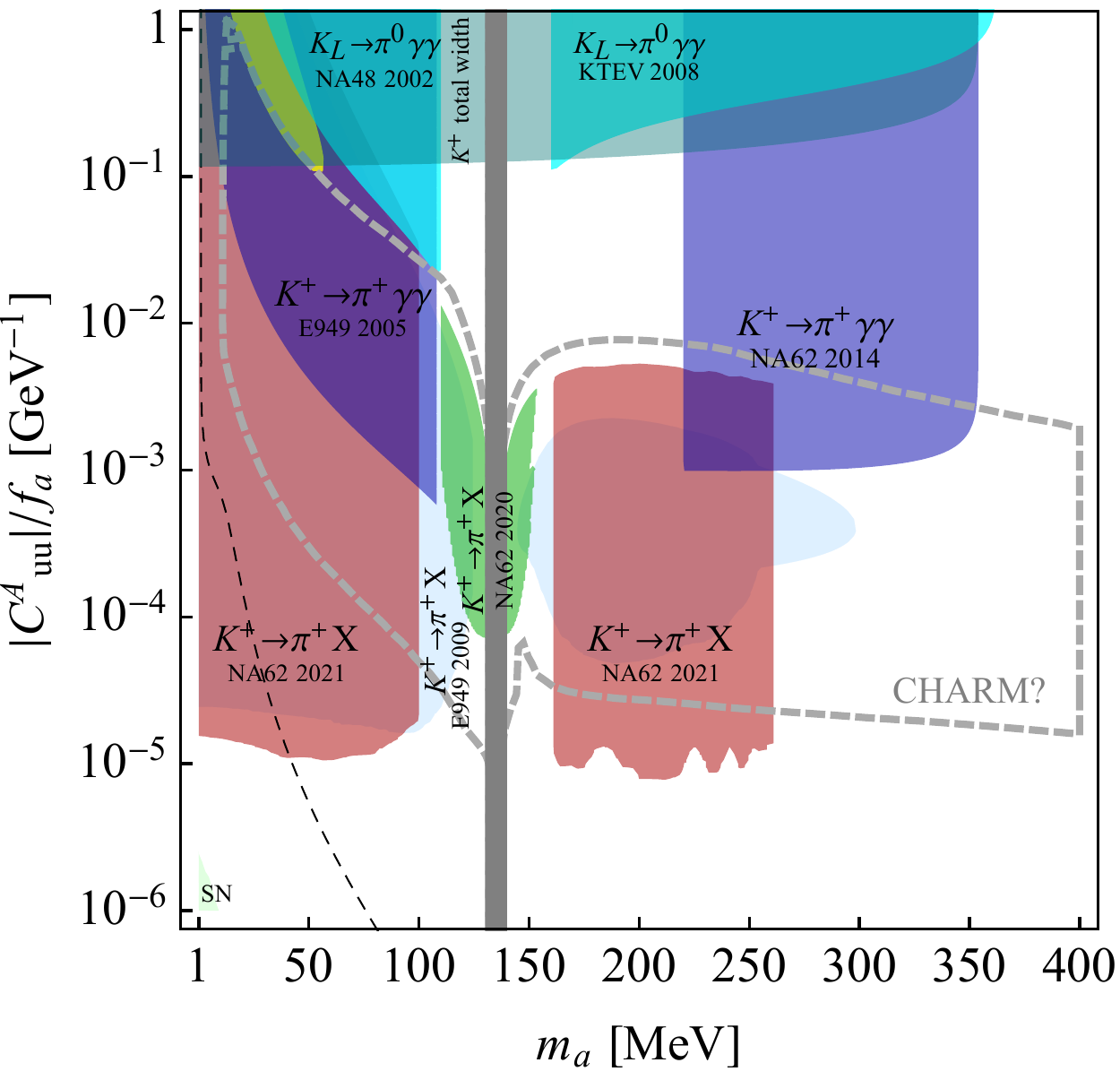}~~~~
\includegraphics[width=3in]{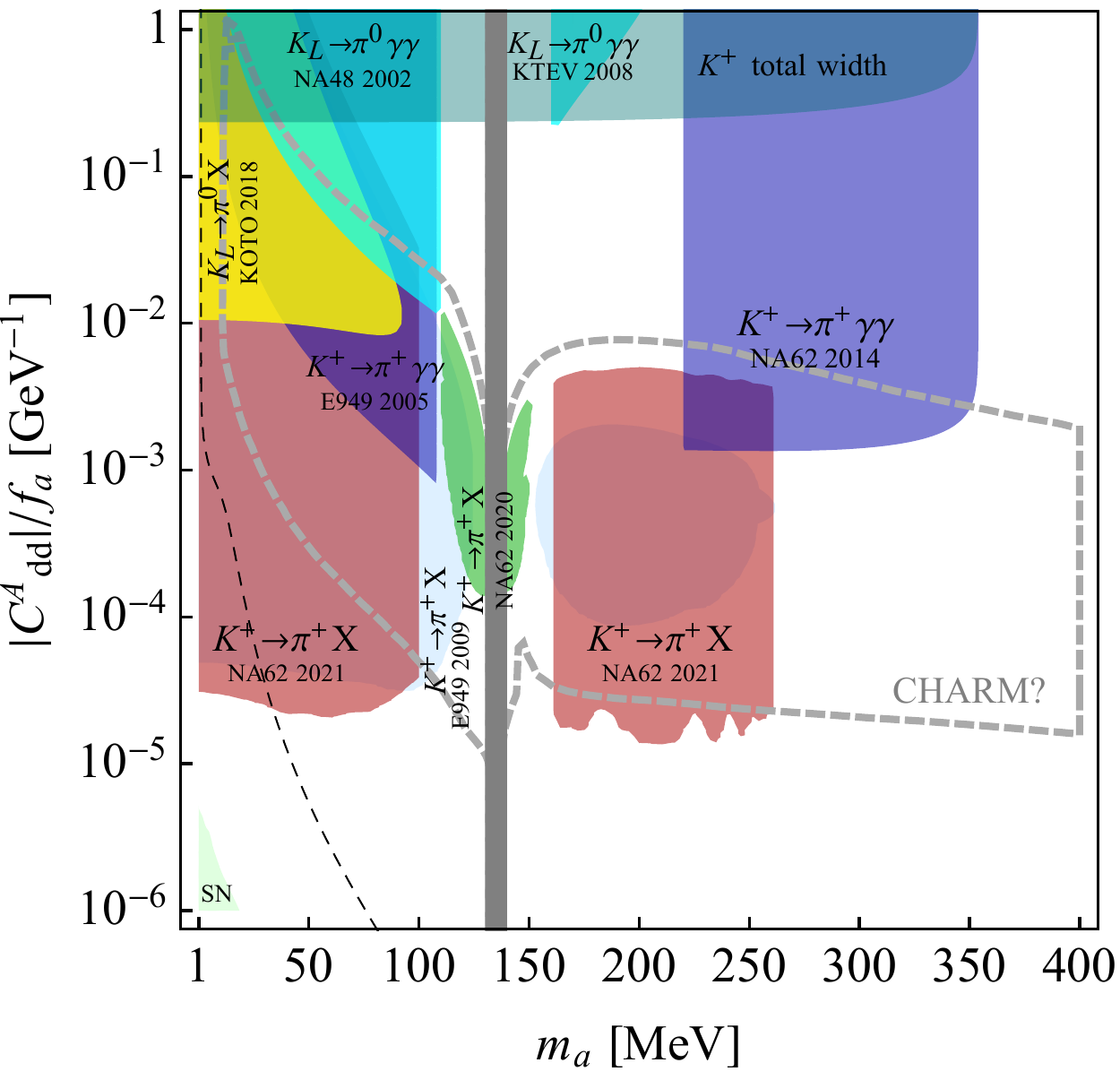}
\caption{
\label{fig:cggcwweeplots}
\small Upper left: Bounds on the ALP that at tree level couples only to gluons and leptons with $C^A_{\ell\ell}=N_3=1$. The coupling to photons is induced at loop level through equation~\eqref{eq:cgamgameff}, allowing the ALP to decay to two photons. Upper right: Bounds on ALP with tree level couplings to only $SU(2)_L$ gauge bosons and leptons,  $C^A_{\ell\ell}=N_2=1$. The ALP decay to photons occurs at tree level through the $N_2$ coupling. Bottom: Bounds on flavor-diagonal pseudoscalar quark couplings of ALP to only the first generation quarks: coupling to up quarks (left) and down quarks (right) with couplings to  loop induced photons. Figure from \cite{Goudzovski:2022vbt}. \label{fig:axionplot2}}
\end{figure}

\paragraph{ALP with several phenomenologically relevant couplings.}
 When experimental searches are performed it is important not to leave any stone unturned.  For instance, non-minimal implementations of PQ symmetry allow for the strong CP problem to be solved by heavier axions than one might naively expect~\cite{Dimopoulos:1979pp,Holdom:1982ex,Holdom:1985vx,Dine:1986bg,Flynn:1987rs,Choi:1988sy,Rubakov:1997vp,Choi:1998ep,Berezhiani:2000gh,Choi:2003wr,Hook:2014cda,Fukuda:2015ana,Dimopoulos:2016lvn,Agrawal:2017ksf,Agrawal:2017eqm,Gaillard:2018xgk,Hook:2019qoh,Gherghetta:2020keg,Kitano:2021fdl,Gupta:2020vxb}.  In Fig. \ref{fig:cggcwweeplots} we show an example of ALP constraints where the mass of the ALP is treated as a free parameter, and several ALP couplings in \eqref{L_ALP} are nonzero. Fig. \ref{fig:cggcwweeplots} (upper left) shows the constraints in the case of a leptophilic ALP with universal couplings to leptons $C_{\ell \ell}^A=1$ and that also couples to gluons with $N_3=1$.  This leads to stringent constraints, with almost all parameter space below $f_a/N_3\sim 10$ TeV excluded by searches for rare decays into ALP either decaying to electrons, muons or photons. A qualitatively different situation is obtained for the case of leptophilic ALP coupling to electroweak gauge bosons, $N_2=C_{\ell\ell}^A=1$, shown in Fig. \ref{fig:cggcwweeplots} (upper right). The bounds, especially for lighter ALP, are then significantly relaxed. 
 
 \paragraph{ALP with flavor non-universal couplings.}
 Another example of non-trivial ALP couplings is the possibility of flavor aligned but non-universal ALP couplings. Fig. \ref{fig:cggcwweeplots}, bottom two panels, show the constraints on ALP that only couples to first generation up (bottom left) or down (bottom right) quarks. Since in this case the flavor violation only occurs due to the SM flavor violating sources, the exchanges of $W$ bosons, the constraints from $K\to \pi a$ decays on $f_a$ are much less severe than if such decays were due to flavor violating ALP couplings in the UV (cf. constraints from FNCN transitions in Fig. \ref{fig:axionplot}).
 
\paragraph{Leptophilic ALP with flavor violating couplings.}
If the spontaneously broken U(1) is acting on the lepton sector, this will result in a leptophilic  ALP with predominant SM couplings to leptons. A prime example of such an ALP is the Majoron, a pNGB of a spontaneously broken global lepton number~\cite{Chikashige:1980ui, Schechter:1981cv}, with other examples such as lepton flavor violating QCD axion, lepton flavor violating axiflavon and leptonic familon also discussed in the literature \cite{Calibbi:2020jvd}. Such ALPs with flavor violating couplings could well be first discovered in the experiments that are searching for flavor changing neutral current processes with leptons. 

\begin{figure}[t]
\centering
\includegraphics[width=5in]{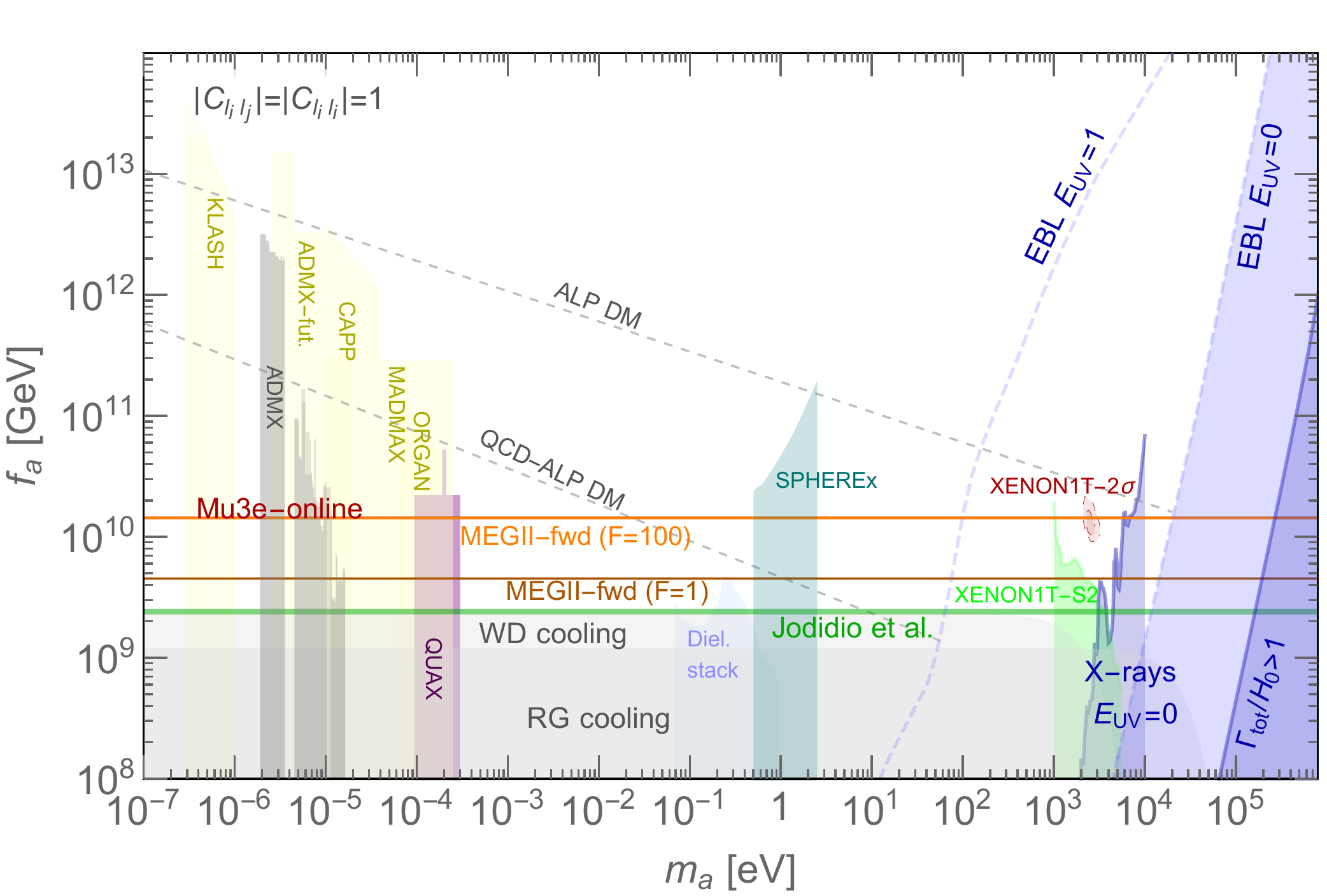}~~~~
\caption{The impact of present and future $\mu\to e a$  searches compared to other light ALP DM searches, taking $E_{\text{UV}}=1,0$ as two representative examples, see main text for details. Figure from \cite{Calibbi:2020jvd}. 
\label{fig:leptpnALP}
\small }
\end{figure}

Fig. \ref{fig:leptpnALP} shows constraints on a leptophilic ALP with flavor violating couplings, assuming that it is also the dark matter.  The  green solid line shows the current best bound on the isotropic LFV ALP~\cite{Jodidio:1986mz}, the (dark) orange thin line gives the MEGII-fwd projection assuming $F=100$ focusing enhancement (no focusing) \cite{Calibbi:2020jvd}. The {dark red} line (overlapping with the orange thin line) shows the sensitivity of Mu3e-online analysis~\cite{Perrevoort:2018okj}. In the blue region enclosed by the { solid blue} line the ALP decays within the present Hubble time, while the region to the right of the { dashed blue} line is excluded by the extragalactic diffuse background light measurements for two different benchmark values of ALP-photon couplings, $E_{UV}=0,1$, where $E_{\rm UV}=(N_1+N_2)/2$ in \eqref{eq:cgamgameff}, as well as the X-rays constraints for $E_{\text{UV}}=0$~\cite{Boyarsky:2006hr,Figueroa-Feliciano:2015gwa}. The red blob indicates where ALP DM could explain the XENON1T anomaly~\cite{Bloch:2020uzh}. The { dashed gray lines} denote two scenarios where the observed DM relic abundance is due to ALPs  produced through the misalignment mechanism, either with temperature independent ALP mass, or with temperature dependence parametrically  similar to the one for the QCD axion. The { gray shaded} regions are excluded by the star cooling bounds, and the ADMX data~\cite{Braine:2019fqb,Boutan:2018uoc,Du:2018uak}. The light green region is excluded by the S2 only analysis of XENON1T~\cite{XENON:2019gfn} and Panda-X~\cite{Fu:2017lfc}. The { purple shaded} region shows the future reach of axion-magnon conversion experiments such as QUAX~\cite{Barbieri:1985cp,Barbieri:2016vwg,Chigusa:2020gfs}. Regarding the coupling to photons, the { cyan band} shows the future sensitivity of SPHEREx estimated in Ref.~\cite{Creque-Sarbinowski:2018ebl}, assuming ALP decay exclusively to two photons, while the { yellow bands} show the future sensitivities of resonant microwave cavities such as ADMX~\cite{Shokair:2014rna}, CAPP~\cite{Petrakou:2017epq}, KLASH~\cite{Gatti:2018ojx}, and ORGAN~\cite{McAllister:2017lkb}, dielectric haloscopes such as MADMAX~\cite{TheMADMAXWorkingGroup:2016hpc} and the reach of the dielectric stack proposal~\cite{Baryakhtar:2018doz} is shown with { light blue}.

\section{Mapping of models to Experimental Approaches}
\label{sec:experiments}
Coverage towards unique final states is critical when considering the experimental reach for extended dark sectors and anomaly-motivated models. The models in this paper go beyond the conventional models and highlight specific final states and parameter choices that differ from the conventional dark sector benchmarks. By going beyond the conventional models, we can introduce novel final states, explain anomalous features in the data, and explore fundamental theoretical problems. While the full scope of the dark sector experiments is not covered here, we would like to highlight the \emph{specific} features that frequently arise when considering extended dark sector models. 

A few specific signatures are emphasized when searching for extended dark sector models. In particular, there is an enhanced rate of  
\begin{itemize}
    \item Long Lived particles,
    \item Increase in the importance of heavy flavor interactions,
    \item an increase in the importance of non-resonant processes,
    \item and the use of primary and secondary muon beams.
\end{itemize}
When extending the scope of dark sector models, each of the above features deviates differently with respect to the benchmark model signatures. 

For example, the addition of inelastic dark matter production enables the possibility for dark sector particles to decay with a lifetime that scales with the mass splitting $\Delta$ between the degenerate particle states. This essentially makes the lifetime a free parameter with limited constraints. As a consequence, to increase sensitivity to inelastic dark matter and related models, experiments need to be able to detect long lived particle decays over a broad range of lifetimes ranging from $c\tau\sim1$\,mm to $c\tau\sim10^5$\,km. The range of lifetimes is enlarged when compared to the minimal dark sector models, where the standard model couplings determine the lifetime. The increased scope of possible lifetimes motivates detector designs that accommodate this significant variation. Consequently, different experimental enhancements are needed to contend with other features of extended dark sector models. Below, we highlight the critical elements required to ensure sensitivity to extended dark sector models. 

{\bf Long Lived Particles:} As with inelastic dark matter, a broad range of models introduce unstable dark sector particles with long lifetimes. Additionally, weakly coupled mediators that are lighter than dark sector particles can also have long lifetimes because they are kinematically restricted to decay to standard model particles and have a weak coupling. The range of possible lifetimes can be extensive and depends on the model. However, to avoid impacts on big bang nucleosynthesis, lifetimes less than one second ($c\tau \lesssim 3\times10^{8}$~m) are required. 

The ability to detect long lived particles requires an intense beam and the ability to shield decay remnants from the intense beam. As a result, experiments that can cover a range of different lifetimes are essential to improve sensitivity to long lived particles. Lastly, when the lifetime becomes very long and, as a result, the probability for decay within the fiducial volume is small, the final state ends up being bounded by missing energy searches. 

Various experimental searches for long lived particles are being considered in the next decade. These include the DarkQuest experiment~\cite{Apyan:2022tsd}, Codex-b\cite{Aielli:2022awh}, the GAZELLE experiment at Belle-II~\cite{Dreyer:2021aqd}, the Heavy Photon Search~\cite{Baltzell:2022rpd}, the forward physics facility~\cite{Anchordoqui:2021ghd}, and, for missing energy searches, the LDMX experiment~\cite{Akesson:2022vza}. Lastly, in the low dark matter mass region, neutrino experiments including JSNS$^{2}$ \cite{Endo:2022imj}, and MicroBoone, as well as PIP-II~\cite{Toups:2022yxs}, are relevant.

The beam dump approach for each experiment varies. The JSNS$^{2}$ \cite{JSNS2:2021hyk}, MicroBoone~\cite{MicroBooNE:2021zai,MicroBooNE:2021ewq}, and Forward Physics Facility \cite{Feng:2022inv} rely on large amounts of shielding, to limit the beam. The DarkQuest experiment uses a strong magnetic field to sweep away the beam dump remnants, allowing for shorter lifetimes to be probed. The HPS \cite{Baltzell:2022rpd} and NA62 \cite{NA62:2021zjw} experiments utilize long decay volumes and less intense beams. Finally, the LHC experiments rely on a further lower intensity beam and elaborate instrumentation to probe the full parameter space.  

{\bf Heavy Flavor signatures:} Kaon decay experiments and muon decay experiments are potent probes of axion and other flavor misaligned physics scenarios. Flavor violation can induce rare kaon and muon decays that have the distinct ability to probe flavor misalignment at very high scales and/or weak couplings. As a consequence, experiments capable of probing a large variety of rare kaon decays, such as NA62 \cite{NA62:2021zjw}, and the KOTO \cite{KOTO:2020prk} experiment are critical towards putting relevant and critical bounds on flavor violating models. For maximal flavor violation, these experiments can probe the QCD axion through $K\to\pi a$ decays. Furthermore, anomalous muon decay experiments, in particular $\mu\rightarrow e\gamma$, such as MEG and MEG-II, can put bounds directly on ALP lepton couplings. 

{\bf Non-resonant processes:} More complex scenarios such as that present in SIMP models can lead to signatures that involve mixtures of standard model particles and dark sector decays, a signature that may not lead to clear resonances in experimental searches. Consequently, searches for such final states often have to rely on the complexity of the decay topology to separate the signal from the background. As a result, the need to reconstruct various final state decays and the ability to look for multiple, differently displaced particles can be essential elements that are important to resolve these complex topologies. Experiments that are sensitive to long lived complex topologies can reconstruct a variety of final states and lifetimes. These include the SDarkQuest experiment~\cite{Apyan:2022tsd}, and the Belle-II experiment, in conjunction with GAZELLE~\cite{Dreyer:2021aqd}. Additionally, LHC long lived searches such as FASER, CODEX-b, and MATHUSLA are capable of dark sector searches~\cite{Feng:2022inv,Aielli:2022awh,MATHUSLA:2022sze}. 

{\bf Primary and Secondary Muon Beams:}  The importance of $(g-2)_{\mu}$ has highlighted the need to look for models with enhanced interactions to 2nd and 3rd generations. This is a major motivating factor for enhanced use of muon beams and enhanced detection of final states from the heavy flavor. Muon beams, in particular, are the most powerful ways to probe new physics mediator coupling to muons. These muon beams can be produced either through a dedicated muon beam or as secondary muons from an intense proton or electron beam. Currently, muon beams can be created at the SPS facility and are planned to be used with NA64~\cite{Sieber:2021fue} and the MuOnE experiment \cite{Masiero:2020vxk}. Secondary muon beams can be produced through an electron beam with the BDX experiment \cite{BDX:2019afh}, with the proton beam at the SeaQuest/DarkQuest experiment, and through muons produced at the LHC. 

For $(g-2)_{\mu}$, the dominant bounds tend to come from muon beam dump experiments at low mass and decays from upsilon or other heavy resonances at high mass. In the next decade, there is potential to probe well motivated $(g-2)_{\mu}$ models with muon beams from NA64~\cite{Sieber:2021fue},  Belle-II~\cite{Asner:2022axe}, and at high mass, the LHC~\cite{CMS:2018yxg}. Furthermore, heavy flavor and $(g-2)_{\mu}$ motivate research towards the construction of an intense muon beam where muonic dark sectors can be probed directly through experiments such as $M^{3}$~\cite{Kahn:2018cqs}. Future muon beam facilities that are being considered could have a significant impact on probing flavor specific signatures~\cite{Arrington:2022pon}. 

\section{Conclusions}
Non-minimal (or ``complex'') dark matter scenarios can readily help resolve a range of important physics problems. In particular, they can explain certain experimental anomalies, address core theoretical issues with the Standard Model, and allow for novel signatures. Such extended dark sector models can explain recent data anomalies, including the observation of muon $(g-2)_{\mu}$, the asymmetry in the lepton flavor final states of $B$ hadron decays, and the Xenon 1T excess. Moreover, these extended scenarios can help explain core fundamental theoretical anomalies such as the QCD axion. Extended dark sector models can allow for new mechanisms to produce dark matter in the early universe, such as that of $3\to 2$ processes in SIMP models. Lastly, extended dark sector models lead to new possible final states at collider experiments. 

Specific detector features are favored for specific non-minimal sector final states to search for extended dark sectors. In many models, the presence of long lived signatures and partially visible final states appear as a result of adding additional particles with the dark sector model. Amongst the examples considered here, we observe that both inelastic dark matter and SIMPs have semi-visible and long-lived signatures. Furthermore, flavor violating decays from rare decays from Kaon and muon beams can lead to powerful constraints on flavor violating dark sector decays, which critically probe the properties of the QCD axion. Finally, muon beams and secondary muon beams from proton and electron beam dumps are critical probes of modified dark sector models that target heavy flavor anomalies and can explain the current $(g-2)_{\mu}$ anomaly.

Over the next decade, there is an exciting opportunity to powerfully test the best-motivated dark matter related explanations for the $(g-2)_{\mu}$ and QCD anomalies, provided that an appropriate set of proposed and existing dark sector searches are carried out to completion. To ensure this success, we must exploit the existing capabilities of large multi-purpose detectors, such as Belle-II and LHCb, and future upgrades. Furthermore, we need to invest in specialized, small-scale experiments such as the DarkQuest experiment, LDMX, and HPS and the facilities that drive these experiments. Moreover, we should consider investing in muon facilities that can definitively probe $(g-2)_{\mu}$ anomaly and Kaon facilities that can put substantial constraints on the QCD axion. Finally, we aim to support the theory community further to ensure a robust and diverse exploration of dark sector physics. 





\vspace{0.2in}

{\bf Acknowledgements.} 
J.Z. acknowledges support in part by the DOE grant de-sc0011784 and NSF OAC-2103889.  P.H. acknowledges support by DOE grant de-sc0021943.

\bibliographystyle{JHEP}

\bibliography{BG3}

\end{document}